\documentclass[aps,prd,showpacs,preprint,superscriptaddress]{revtex4}
\usepackage{amssymb}
\usepackage{amsmath}
\usepackage{graphicx}
\usepackage{multirow}
\usepackage{slashed}
\usepackage{mathrsfs}
\usepackage{subfigure}
\usepackage{float}

\begin{document}

\title{Signature of same-sign top pair production mediated by a
 nonuniversal $Z^\prime$ with QCD next-to-leading order accuracy at the LHC}

\author{Bo Hua Li}
\affiliation{Department of Physics and State Key Laboratory
of Nuclear Physics and Technology, Peking University,
Beijing 100871, China}

\author{Chong Sheng Li}
\email{csli@pku.edu.cn}
\affiliation{Department of Physics and State Key Laboratory
of Nuclear Physics and Technology, Peking University,
Beijing 100871, China}
\affiliation{Center for High Energy Physics,
Peking University, Beijing 100871, China}

\author{Hai Tao Li}
\affiliation{Department of Physics and State Key Laboratory
of Nuclear Physics and Technology, Peking University,
Beijing 100871, China}

\author{Yong Chuan Zhan}
\affiliation{Department of Physics and State Key Laboratory
of Nuclear Physics and Technology, Peking University,
Beijing 100871, China}

\author{Yue Zhang}
\affiliation{Department of Physics and State Key Laboratory
of Nuclear Physics and Technology, Peking University,
Beijing 100871, China}

\author{Jian Wang}
 \affiliation{Department of Physics and
State Key Laboratory of Nuclear Physics and Technology, Peking
University, Beijing 100871, China}

\date{\today}


\begin{abstract}
We present a detailed study of the
same-sign top pair production mediated by a nonuniversal $Z^\prime$ including production
and decay at the Large Hadron Collider (LHC) at the  QCD next-to-leading order (NLO) level, using the narrow width approximation
 and helicity amplitudes method. We find that
the QCD NLO corrections can loosen the constraint on the model parameters and
reduce the dependence of the total cross sections on the factorization
scale significantly. We also study the signature and backgrounds of the process at the NLO level.
In order to suppress the backgrounds, we further investigate the difference between the production rates of the positively and negatively charged dilepton at the LHC, and find that the same-sign dilepton signal of the new physics
could be discovered more easily. Besides, we also discuss the uncertainty from the parton distribution (PDF) at the NLO level.

\end{abstract}

\maketitle

\section{introduction}\label{s1}
Up to now, the top quark is the heaviest fundamental particle with a mass close
to the electroweak (EW) symmetry breaking scale.
Thus it would be more sensitive to the new physics beyond the standard model (SM). One way to study the
new physics in the top quark sector is via the anomalous flavor-changing neutral-current (FCNC) couplings.
Within the SM, the FCNC couplings  are absent at tree level and occur through loop diagrams, which are further suppressed by the Glashow-Iliopoulos-Maiani mechanism~\cite{Glashow:1970gm}.
However, in some new physics models, such as two Higgs doublet model~\cite{Cheng:1987rs}, supersymmetric models~\cite{Li:1993mg}, extra dimensions models~\cite{Davoudiasl:2001uj},
little Higgs models~\cite{HongSheng:2007ve}, and the $Z'$ model~\cite{Jung:2009jz,Gresham:2011dg}, the FCNC processes can occur at tree level, which may enhance the cross section to observable level.

Due to the small SM backgrounds for the same-sign dilepton final state, the same-sign top pair production is a good channel to study the FCNC couplings. There are already a lot of articles~\cite{Berger:2011ua,Gupta:2010wx,Larios:2003jq,BarShalom:2007pw,Gao:2008vv,Martin:2008aw,Zhang:2010kr}
discussing the same-sign top pair production process, 
and in most of there, the process is induced by the FCNC couplings.
In general, these FCNC couplings can be divided into several types. One of the most important types involves a massive
colorless vector boson, i.e., $Z^\prime$, which is the gauge boson associated with additional $U(1)$ symmetries.
The renormalizable FCNC interaction can be generally written as follows~\cite{Jung:2009jz}
\begin{equation}\label{L}
  \mathcal{L}= \bar{u}\gamma^\mu (C_R P_R + C_L P_L)t Z_\mu^\prime + h.c.,
\end{equation}
where $C_R$ and $C_L$ are the coupling strength and $P_R$ and $P_L$ are the projection operators. Because of
the constraint from the $B_d-\bar{B_d}$ mixing~\cite{Cao:2010zb},
only the right-handed coupling is considered below.

Due to the high energy and luminosity at the Large Hadron Collider (LHC),
abundant top quark events will be produced, so it is a good chance to
investigate the same-sign top pair production process. In fact, from the measurements of the total cross sections,
the CMS Collaboration has already set an upper limit on the parameters of the nonuniversal $Z^\prime$~\cite{Chatrchyan:2011dk,Chatrchyan:2012sa}, and at the low $Z^\prime$ mass region the ATLAS Collaboration also gave a strong constraint~\cite{Aad:2012cg}. However, both of these results are based on the leading-order (LO) calculations, which suffer from large-scale uncertainties and cannot match the expected experimental accuracy at hadron colliders. On the other hand, the backgrounds of the process can be further suppressed if we choose a more proper observable, as we will discuss below.
In this paper, we present the complete QCD next-to-leading order (NLO) corrections to the
same-sign top pair production and decay (lepton channel) mediated by the nonuniversal
$Z^\prime$ at the LHC and also investigate the signal and backgrounds of the process at the QCD NLO level. Note that
associated production of a top quark and a $Z^\prime$ boson via this coupling was studied in Ref.~\cite{Adelman:2012py} at NLO.

The arrangement of the paper is as follows. In Sec.~\ref{s2}, we describe the
method used in our calculation. In Sec.~\ref{s3} we
show the LO results for the process. In Sec.~\ref{s4} we present
the details of the NLO calculations for the production and the decay processes.
We give our numerical results in Sec.~\ref{s5},
and Sec.~\ref{s6} is a brief summary.

\section{Narrow width approximation and helicity amplitudes method}\label{s2}
The narrow width approximation (NWA)\cite{Kauer:2007zc,Uhlemann:2008pm} is often used for a resonant
 process when the heavy resonance has a small decay width.
If the resonance is a scalar, the total cross section can be separated into
two parts, i.e., production and decay,
\begin{eqnarray}
\label{eq:nwa1}
\sigma &=& \frac{(2\pi)^7}{2s}\int_{q^2_\text{min}}^{q^2_\text{max}}dq^2\int d\phi_p d\phi_d
|\mathcal{M} _p(q^2)|^2 \left[\left(q^2-M^2\right)^2+(M\Gamma)^2\right]^{-1} |\mathcal{M} _d(q^2)|^2
               \nonumber \\
 &=& \frac{(2\pi)^8}{4s M\Gamma}\int d\phi_p|\mathcal{M} _p(M^2)|^2 \int d\phi_d|\mathcal{M} _d(M^2)|^2,
\end{eqnarray}
where $M$ is the mass of the resonance and $\Gamma$ is the decay width of the resonance. The
$\mathcal{M} _p$, $\mathcal{M}_d$ are the amplitudes of
the resonance production and its decay, respectively.

However, in our case, the resonance is the top quark,
so the process should be separated at the amplitude level.
Since the intermediate top quark is on-shell~\cite{Melnikov:2009dn,Campbell:2004ch}, its propagator can be
written as
\begin{equation}
\frac{\slashed q + m_t}{\left(q^2-m_t^2\right)+i(m_t\Gamma_t)}
=\frac{ u_+\bar u_+ +  u_-\bar u_-}{\left(q^2-m_t^2\right)+i(m_t\Gamma_t)}
\end{equation}
where $u_+$ and $u_-$ denote the top quark spinors with positive and negative helicities, respectively. $m_t$
is the top quark mass, and $\Gamma_{t}$ is the total decay width of the top quark. Then
 Eq.~(\ref{eq:nwa1}) would be changed to the following form
\begin{eqnarray}
\label{eq:nwa2}
\sigma &=&
\frac{(2\pi)^8}{4s M\Gamma}\bigg(\int d\phi_p |\mathcal{M} _p^+|^2 \int d\phi_d|\mathcal{M} _d^+|^2
+\int d\phi_p |\mathcal{M} _p^-|^2  \int d \phi_d  |\mathcal{M} _d^-|^2\bigg),
\end{eqnarray}
where $\mathcal{M} _p^\pm$, $\mathcal{M} _d^\pm$ are the helicity amplitudes of
 the same-sign top pair production and the top quark decay,
respectively. It should be note that if
$\int d\phi_p|\mathcal{M} _p^+|^2 = \int d\phi_p|\mathcal{M} _p^-|^2$ or
$\int d\phi_d|\mathcal{M} _d^+|^2 = \int d\phi_d|\mathcal{M} _d^-|^2$, after
a polarization average factor $\frac{1}{2}$ is taken into account,
 Eq.~(\ref{eq:nwa2}) will be equivalent to Eq.~(\ref{eq:nwa1}).

We adopt the helicity amplitude method in our calculation. The massless spinors are denoted as \cite{Badger:2010mg}
\begin{equation}
|i^{\pm}\rangle\equiv u_{\pm}\left(k_{i}\right)=v_{\mp}\left(k_{i}\right),\quad\langle i^{\pm}|\equiv\overline{u}_{\pm}\left(k_{i}\right)=\overline{v}_{\mp}\left(k_{i}\right),
\end{equation}
and massive spinors can be written as
\begin{eqnarray}
&&u_{\pm}\left(p,M;\eta,p^{\flat}\right)=\frac{\left(\slashed{p}+M\right)|\eta^{\mp}\rangle}{\langle p^{\flat\pm}|\eta^{\mp}\rangle},\quad\overline{u}_{\pm}\left(p,M;\eta,p^{\flat}\right)=\frac{\langle\eta^{\mp}|\left(\slashed{p}+M\right)}{\langle\eta^{\mp}|p^{\flat\pm}\rangle},\nonumber\\
&&v_{\pm}\left(p,M;\eta,p^{\flat}\right)=\frac{\left(\slashed{p}-M\right)|\eta^{\pm}\rangle}{\langle p^{\flat\mp}|\eta^{\pm}\rangle},\quad\overline{v}_{\pm}\left(p,M;\eta,p^{\flat}\right)=\frac{\langle\eta^{\pm}|\left(\slashed{p}-M\right)}{\langle\eta^{\pm}|p^{\flat\mp}\rangle},
\end{eqnarray}
where $p^{\flat}$ and $\eta$ are two massless momenta, which fulfill the following
conditions
\begin{equation}
p=p^{\ensuremath{\flat}}+\frac{M^{2}}{2p\cdot\eta}\eta,\quad p^{2}=M^{2},\quad\left(p^{\flat}\right)^{2}=\eta^{2}=0.
\end{equation}

Since the helicity of massive spinors have the following relations
\begin{equation}
\frac{\langle p^{\flat\mp}|\eta^{\pm}\rangle}{M}\overline{u}_{\pm}\left(p,M;p^{\flat},\eta\right)=\overline{u}_{\mp}\left(p,M;\eta,p^{\flat}\right),\quad\frac{\langle p^{\flat\mp}|\eta^{\pm}\rangle}{M}v_{\pm}\left(p,M;p^{\flat},\eta\right)=v_{\mp}\left(p,M;\eta,p^{\flat}\right),
\end{equation}
we only present results for amplitudes of the same-sign top pair production with a $(...,t^+, t^+)$ helicity
 configuration.

\section{Leading-Order Results}\label{s3}

\begin{figure}[h]
\begin{center}
\scalebox{0.7}{\includegraphics*{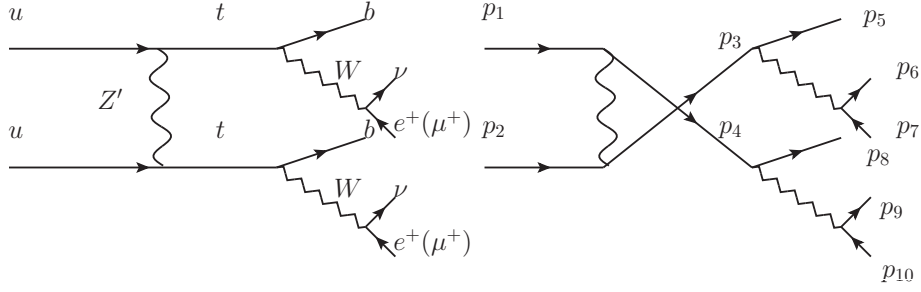}}
\caption[]{\label{tree}The LO Feynman diagrams for the same-sign top pair
production and decay at the leading order via the $Z^{\prime}$ FCNC couplings.}
\end{center}
\end{figure}

At the LHC there is only one subprocess that contributes to the same-sign top pair
production and decay (lepton channel) at the LO via the $Z^{\prime}$ FCNC couplings,
\begin{equation}
u \ u \longrightarrow t \ t \longrightarrow l^+ \nu b \l^+ \nu b.
\end{equation}
The corresponding Feynman diagrams are shown in Fig.~\ref{tree}, where the top quark is on-shell.
Here and below we adopt the unitary gauge.

The LO helicity amplitudes for $t$-channel and $u$-channel same-sign top pair productions
in four dimensions are

\begin{eqnarray}
\mathcal{M}^{++++}_t &=& \frac{2 C_R^2}{(t-M_{Z^{\prime}}^2)
\langle \eta_3|p_{3}^\flat \rangle \langle \eta_4|p_{4}^\flat \rangle}
\{\langle p_2|p_1 \rangle \langle \eta_4|p_1 \rangle [p_1| \slashed p_3 |\eta_3\rangle +
 \langle p_2|p_1 \rangle \langle \eta_4|p_2 \rangle [p_2| \slashed p_3 |\eta_3\rangle
\nonumber \\
&& - m_{t}^2 \langle p_2|p_1 \rangle \langle \eta_4|\eta_3 \rangle +
 \frac{m_{t}^4}{2M_{Z^{\prime}}^2} \langle \eta_3|p_1 \rangle \langle \eta_4|p_2 \rangle
 \},
 \\
 \mathcal{M}^{++++}_u &=& \frac{-2 C_R^2}{(u-M_{Z^{\prime}}^2)
\langle \eta_3|p_{3}^\flat \rangle \langle \eta_4|p_{4}^\flat \rangle}
\{\langle p_2|p_1 \rangle \langle \eta_3|p_1 \rangle [p_1| \slashed p_4 |\eta_4\rangle +
 \langle p_2|p_1 \rangle \langle \eta_3|p_2 \rangle [p_2| \slashed p_4 |\eta_4\rangle
\nonumber \\
&& - m_{t}^2 \langle p_2|p_1 \rangle \langle \eta_3|\eta_4 \rangle +
 \frac{m_{t}^4}{2M_{Z^{\prime}}^2} \langle \eta_4|p_1 \rangle \langle \eta_3|p_2 \rangle
 \},
\end{eqnarray}
and for the top quark decay
\begin{equation}
\mathcal{M}^{+}_{decay}=-\frac{g^2 m_t U_{tb}\langle p_\nu|p_b \rangle [\eta_t|p_l] }
{((p_t-p_b)^2-M_W^2+i M_W \Gamma_W)[\eta_t|p_{t}^\flat]},
\end{equation}
where $M_{Z^{\prime}}$ and $M_W$ are the masses of the $Z^{\prime}$ and $W$ boson,
respectively. $\Gamma_W$ is the decay width of the $W$ boson.
$U_{tb}$ is the CKM matrix element. $g$ is the weak coupling. $t$ and
$u$ are the Mandelstam variables, which are defined as
\begin{equation}
  t=(p_1-p_3)^2,\ \ u=(p_1-p_4)^2,
\end{equation}
and $p_t$, $p_b$, $p_\nu$, $p_l$ are the four momentum of the top quark, bottom quark,
neutrino and positron(anti-muon).

At the parton level, after the phase space integration, the LO cross sections are given by
\begin{equation}
\hat \sigma^{++}_{B}=\hat \sigma^{++}_{t}+\hat \sigma^{++}_{u}+\hat \sigma^{++}_{tu},
\end{equation}
where
\begin{eqnarray}
\nonumber \hat \sigma^{++}_{t}&=&\frac{1}{2 s}\int d\Gamma_2
\overline{\sum} |\mathcal{M}^{++++}_t|^2,    \\
\nonumber \hat \sigma^{++}_{u}&=&\frac{1}{2 s}\int d\Gamma_2
\overline{\sum} |\mathcal{M}^{++++}_u|^2,    \\
\nonumber \hat \sigma^{++}_{tu}&=&-\frac{1}{ s}\int d\Gamma_2
\overline{\sum} Re(\mathcal{M}^{++++}_t \mathcal{M}^{++++*}_u).
\end{eqnarray}
The $\overline{\sum}$ means the colors of the final-state particles have been summed
over and the colors and the spins of the initial-state particles have been averaged over.
The top quark decay width for lepton channel at LO is
\begin{equation}
\Gamma_{t,l}^{+}=\frac{1}{2 m_t}\int d\Gamma_3
\overline{\sum} |\mathcal{M}^{+}_{decay}|^2.
\end{equation}
The LO total cross section at hadron colliders is obtained by
convoluting the partonic cross section with the parton distribution
functions (PDFs) $G_{i/P}$ for the proton
\begin{equation}
\sigma^B=\sum_{ab}\int dx_1
dx_2\left[G_{a/P_1}(x_1,\mu_f)G_{b/P_2}(x_2,\mu_f)(\sum_{i,j=+,-}\hat \sigma^{ij}_{B}\frac{\Gamma_{t,l}^{i}~\Gamma_{t,l}^{j}}{\Gamma_{t}^2})_{ab}\right],
\end{equation}
where $\mu_f$ is the factorization scale.

\section{QCD NLO Corrections}\label{s4}

\begin{figure}[h]
      \begin{center}
     \scalebox{0.5}{\includegraphics*{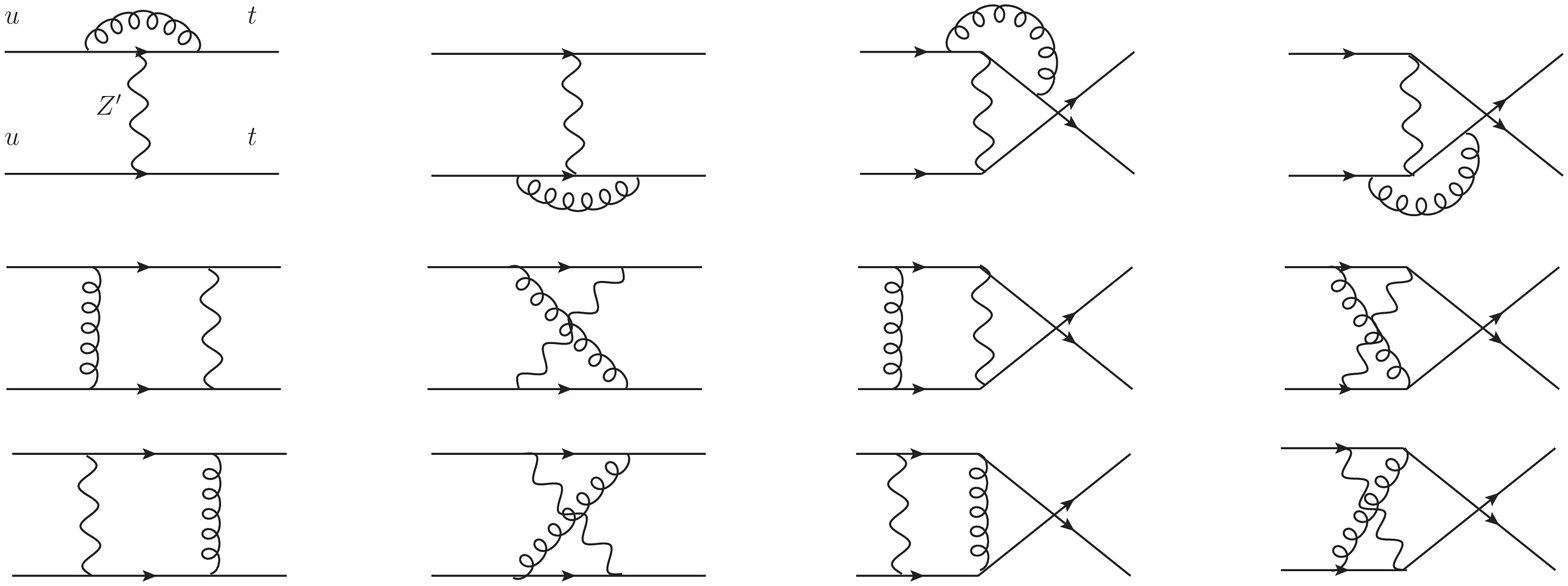}}
      \end{center}
  \caption[]{\label{loop}1-loop Feynman diagrams for the same sign top
pair production via the $Z^{\prime}$ FCNC couplings.}
\end{figure}

\begin{figure}[h]
      \begin{center}
     \scalebox{0.5}{\includegraphics*{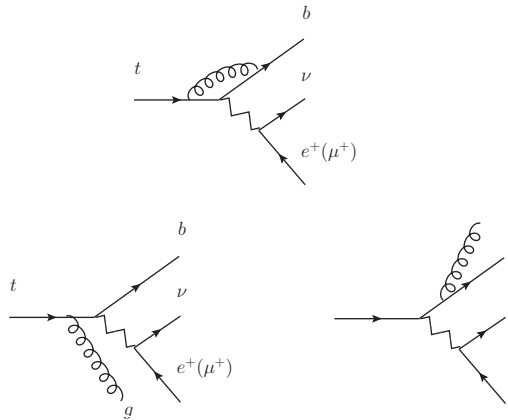}}
      \end{center}
  \caption[]{\label{topdecay} Feynman diagrams for the top decay at the NLO level.}
\end{figure}

The Feynman diagrams for the QCD NLO
corrections to the same-sign top pair production and decay are
shown in Figs.~\ref{loop}~\ref{topdecay}, which include both the
virtual and the real corrections.
The interface between the production
and the decay process, at the NLO level, has been neglected,
because their contributions are suppressed by $O(\frac{\Gamma_{t}}{m_t})$ ~\cite{Fadin:1993kt,Fadin:1993dz,Melnikov:1993np}.

We use the four-dimension helicity (FDH) scheme~\cite{Bern:2002zk} in
$n=4-2\epsilon$ dimensions to regularize all the divergences.
Moreover, for the real corrections, the two cutoff phase space slicing
method~\cite{Harris:2001sx} has been used to separate the infrared (IR) divergences.
\subsection{Virtual corrections}
The squared amplitudes of the virtual corrections are
\begin{equation}\label{eqg5}
\overline{|\mathcal{M}|^2}_{1-loop}=\overline{\sum}2Re(\mathcal{M}^{loop}\mathcal{M}^{B*})
+\overline{\sum}2Re(\mathcal{M}^{con}\mathcal{M}^{B*}),
\end{equation}
where $\mathcal{M}^{loop}$ are the amplitudes for the loop
diagrams, and $\mathcal{M}^{con}$
denotes the corresponding counterterms. Here and below the $\overline{\sum}$ means the colors and spins of the final-state particles have been summed
over and the colors and spins of the initial-state particles have been averaged over.

The virtual corrections contain both UV and IR divergences. Since the
process is induced by the electroweak-type FCNC couplings, the UV
divergences can be cancelled by only introducing the following counterterms:
\begin{eqnarray}
\delta Z_q & = & -\frac{\alpha_s}{3\pi}C_{\epsilon}\left\{ \frac{1}{\epsilon_{\text{UV}}}
-\frac{1}{\epsilon_{\text{IR}}} \right\},
\\
\delta Z_t & = & -\frac{\alpha_s}{3\pi}C_{\epsilon}\left\{ \frac{1}{\epsilon_{\text{UV}}}
+\frac{2}{\epsilon_{\text{IR}}}+5\right\},
\end{eqnarray}
where $C_{\epsilon}=\Gamma(1+\epsilon)[(4\pi\mu_r^2)/m_t^2]
^{\epsilon}$.
We define all the renormalization constants using
the on-shell subtraction scheme.

In Eq.~(\ref{eqg5}), all the UV divergences are canceled out, leaving the IR divergences and the finite
terms. For the top quark production process, the IR divergence of the virtual
corrections at the parton level can be factorized as~\cite{Denner:1991kt,Ellis:2007qk}
\begin{eqnarray}\label{eq1}
 \hat\sigma^{V(IR)}_{pro}&=& -\frac{\alpha_s C_\epsilon C_F}{\pi}\left\{    \right.
(\frac{1}{\epsilon^2}+(\frac{5}{2}-2 a_4)\frac{1}{\epsilon})\hat \sigma_t)+
(\frac{1}{\epsilon^2}+(\frac{5}{2}-2 a_3)\frac{1}{\epsilon})\hat \sigma_u)-
\nonumber \\ &&
(\frac{1}{\epsilon^2}+(a_1+\frac{a_2 \beta }{2 \sqrt{1-\beta }}-
\frac{a_2}{\sqrt{1-\beta }}-2 a_3-2 a_4+\frac{5}{2})\frac{1}{\epsilon})\hat \sigma_{tu})
 \left. \right\},
\end{eqnarray}
where
\begin{eqnarray}
C_F &=& \frac{4}{3}, \quad \beta=\frac{4 m_{t}^2}{s}, \nonumber \\
a_1 &=& \ln(\frac{s}{m_{t}^2}),
\quad  a_2 = \ln \left(\frac{1-\sqrt{1-\beta }}{1+\sqrt{1-\beta }}\right),\nonumber \\
a_3 &=& \ln(\frac{m_{t}^2-u}{m_{t}^2}),
\quad a_4 = \ln(\frac{m_{t}^2-t}{m_{t}^2}).
\end{eqnarray}

For the top quark decay process, the IR divergence of the virtual
corrections to the total cross section can be factorized as~
\begin{eqnarray}\label{eqdecv}
 \Gamma^{V(IR)}_{dec}&=& -\frac{\alpha_s C_\epsilon C_F}{2 \pi}\
(\frac{1}{\epsilon^2}+(5+4 \ln(\frac{m_{t}^2}{2 p_b . p_t}))\frac{1}{2\epsilon})\Gamma_{t,l},
\end{eqnarray}
where $p_t$ and $p_b$ are the top quark and bottom quark momentums, respectively.
In order to cancel these divergences, we need to extract the
IR divergences in the real corrections, which will be shown in the following subsection.

\subsection{Real corrections}

\begin{figure}[h]
      \begin{center}
     \scalebox{0.5}{\includegraphics*{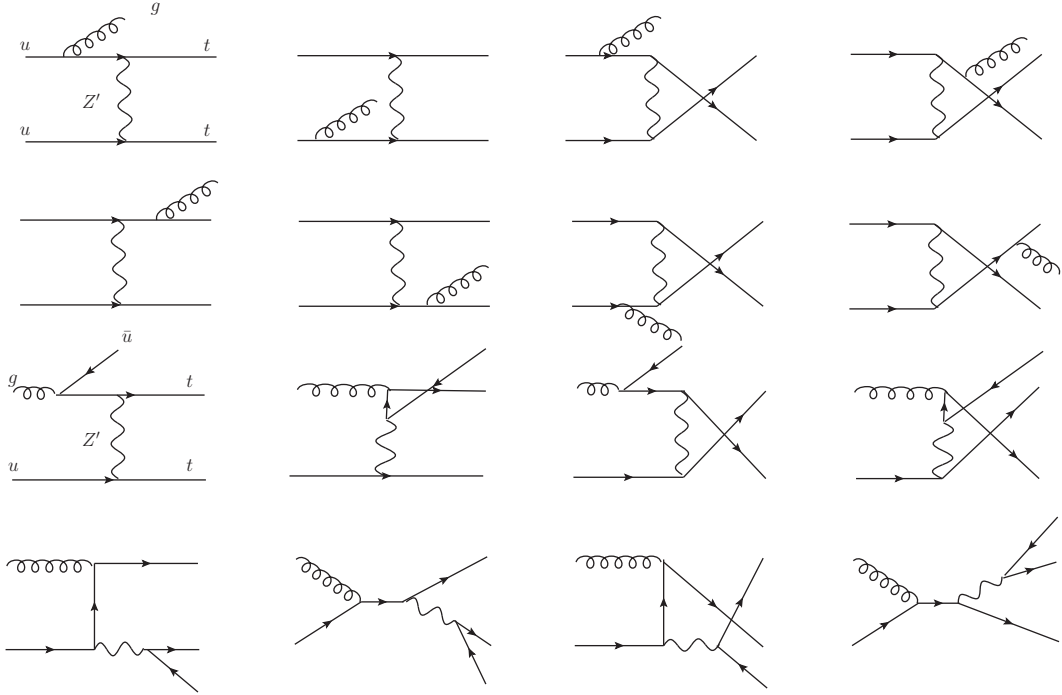}}
      \end{center}
  \caption[]{\label{real}Feynman diagrams of the real corrections for the same sign top
pair production via the $Z^{\prime}$ FCNC couplings.}
\end{figure}

The real corrections consist of the radiations of an
additional gluon $u\ u\longrightarrow t\ t\ g$, or massless anti-quark in the final states,
$\ g\ u\longrightarrow t\ t\ \bar{u} $ as shown in Fig.~\ref{real}. It should be noted that in our NLO
calculations of the process, we include the contributions from the $Z^{\prime}$
on-shell production as the real corrections.

\subsubsection{Real gluon emission}

For real gluon emission, the phase space integration contains both soft
and collinear singularities. We adopt the two cutoff phase space slicing method to
isolate all the IR singularities~\cite{Harris:2001sx}, which introduces two small cutoff parameters $\delta_s$ and
$\delta_c$ to divide the phase space into three parts. The soft
cutoff $\delta_s$ separates the phase space into the soft region
 $E_5\leq\delta_s\sqrt{s}/2$ and the hard region,
\begin{equation}
\hat{\sigma }^R = \hat{\sigma }^{\text{H}}+\hat{\sigma }^S.
\end{equation}
Furthermore, the hard piece can be divided into two subregions by $\delta_c$,
 \begin{equation}
\hat{\sigma }^{\text{H}} = \hat{\sigma }^{\overline{\text{HC}}}+\hat{\sigma }^{\text{HC}}.
\end{equation}

The hard noncollinear part $\hat{\sigma }^{\overline{\text{HC}}}$ is finite
and the phase space integration can be calculated numerically.
For the soft region, in the limit that the energy of the emitted gluon becomes small,
i.e. $E_5\leq \delta_s\sqrt{s}/2$, the amplitude squared
$\overline{\sum}|\mathcal{M}(uu \to t t +g)|^2$ can be
factorized into the Born amplitude squared times eikonal factors
$\Phi_{\text{eik}}^i$
\begin{equation}
\overline{\sum}|\mathcal{M}(uu \to t t +g)|^2
\stackrel{\text{soft}}{\longrightarrow}
(4\pi\alpha_s\mu_r^{2\epsilon}) \overline{\sum}(|\mathcal{M}_t|^{2}
\Phi_{\text{eik}}^a+|\mathcal{M}_u|^{2}\Phi_{\text{eik}}^b
+Re[2\mathcal{M}_t \mathcal{M}_{u}^*]\Phi_{\text{eik}}^c),
\end{equation}
where the eikonal factor is given by
\begin{eqnarray}
\Phi_{eik}^a &=& \frac{C_F}{2} \left\{                      \right.
    \frac{m_t^2-t}{(p_{1}\cdot p_{5})(p_{3}\cdot p_{5})}
  + \frac{m_t^2-t}{(p_{2}\cdot p_{5})(p_{4}\cdot p_{5})}
  - \frac{m_t^2}{(p_{3}\cdot p_{5})^2}
  - \frac{m_t^2}{(p_{4}\cdot p_{5})^2} \left.  \right\},  \\
\Phi_{eik}^b &=& \frac{C_F}{2}\{
    \frac{m_t^2-u}{(p_{1}\cdot p_{5})(p_{4}\cdot p_{5})}
  + \frac{m_t^2-u}{(p_{2}\cdot p_{5})(p_{3}\cdot p_{5})}
  - \frac{m_t^2}{(p_{3}\cdot p_{5})^2}
  - \frac{m_t^2}{(p_{4}\cdot p_{5})^2}\},  \\
\Phi_{eik}^c &=& \frac{C_F}{2}\{\frac{s}{(p_{1}\cdot p_{5})(p_{2}\cdot p_{5})}
  - \frac{m_t^2-t}{(p_{1}\cdot p_{5})(p_{3}\cdot p_{5})}
  - \frac{m_t^2-u}{(p_{1}\cdot p_{5})(p_{4}\cdot p_{5})}
  - \frac{m_t^2-u}{(p_{2}\cdot p_{5})(p_{3}\cdot p_{5})}
\nonumber\\ &-&  \frac{m_t^2-t}{(p_{2}\cdot p_{5})(p_{4}\cdot p_{5})}
  + \frac{m_t^2}{(p_{3}\cdot p_{5})^2}+\frac{m_t^2}{(p_{4}\cdot p_{5})^2}\}.
\end{eqnarray}
Moreover, the three-body phase space
in the soft limit can also be factorized,
\begin{equation}
d\Gamma_3(uu \to tt
+g)\stackrel{\text{soft}}{\longrightarrow}d\Gamma_2(uu \to
tt)dS.
\end{equation}
Here $dS$ is the integration over the phase space of the soft gluon
which is given by
\begin{equation}
dS = \frac{1}{2(2\pi)^{3-2\epsilon}} \int^{\delta_s \sqrt{s}/2}_0
dE_5E_5^{1-2\epsilon}d\Omega_{2-2\epsilon}.
\end{equation}
The parton level cross section in the soft region can be
expressed as
\begin{equation}
\label{sigma:soft1} \hat{\sigma}^S =\sum_i
(4\pi\alpha_s\mu^{2\epsilon}_r)\int
d\Gamma_2\overline{\sum}|\mathcal{M}_{B}^i|^2\int dS \Phi_{\text{eik}}^i.
\end{equation}
Then, after the integration over the soft gluon phase space,
the divergent parts of  Eq.(\ref{sigma:soft1}) become
\begin{eqnarray}
\hat{\sigma}^S &=&
\frac{\alpha_s C_\epsilon C_F}{\pi} \left \{  \right.
(\frac{1}{\epsilon^2}+(1-2 a_4 - 2\ln(\delta_s) )\frac{1}{\epsilon})\hat \sigma_t)+
\nonumber \\ &&
(\frac{1}{\epsilon^2}+(1-2 a_3 - 2\ln(\delta_s))\frac{1}{\epsilon})\hat \sigma_u)-
(\frac{1}{\epsilon^2}+(a_1+\frac{a_2 \beta }{2 \sqrt{1-\beta }}-
\nonumber \\ &&
\frac{a_2}{\sqrt{1-\beta }}-2 a_3-2 a_4+1- 2\ln(\delta_s))\frac{1}{\epsilon})\hat \sigma_{tu})
\left. \right\}.
\end{eqnarray}
In the hard collinear region, $E_5> \delta_s\sqrt{s}/2$ and
$-\delta_c s< t_{i5} < 0$, the emitted hard gluon is collinear to
one of the incoming partons. As a consequence of the factorization
theorem\cite{Collins:1985ue, Bodwin:1984hc} the matrix element
squared for $uu \rightarrow tt +g$ can be factorized into the
product of the Born amplitude squared and the Altarelli-Parisi
splitting function
\begin{equation}
\overline{\sum}|\mathcal{M}(uu \rightarrow tt +
g)|^2\stackrel{\text{collinear}}
{\longrightarrow}(4\pi\alpha_s\mu^{2\epsilon}_r)\overline{\sum}
|\mathcal{M}^B|^2\left(\frac{-2P_{qq}(z)}{zt_{15}}
+\frac{-2P_{qq}(z)}{zt_{25}}\right),
\end{equation}
where $z$ denotes the fraction of the momentum of the incoming
parton carried by $q(g)$, and the unregulated Altarelli-Parisi splitting function
in our case is written explicitly as~\cite{Altarelli:1977zs}
\begin{eqnarray}
P_{qq}(z) &=& C_{F}\Big(\frac{1+z^{2}}{1-z}\Big).
\end{eqnarray}

Moreover, the three-body phase space can also be factorized in the collinear limit.
For example, in the limit $-\delta_c s < t_{15} < 0$, it has the following form
\begin{equation}
d\Gamma_3(uu \rightarrow tt +
g)\stackrel{\text{collinear}}{\longrightarrow}d\Gamma_2(uu
\rightarrow tt; s^{\prime} = zs)
\frac{(4\pi)^{\epsilon}}{16\pi^2\Gamma(1-\epsilon)}dzdt_{15}[-(1-z)t_{15}]^{-\epsilon}.
\end{equation}
Thus, after convoluting with the PDFs, the three-body cross section
in the hard collinear region is given by
\begin{eqnarray}
d\sigma^{HC} & = & d\hat{\sigma}^B
\left[\frac{\alpha_s}{2\pi}\frac{\Gamma(1-\epsilon)}{\Gamma(1-2\epsilon)}
\left(\frac{4\pi\mu^2_r}{s}\right)^{\epsilon}\right](-\frac{1}{\epsilon})
\delta_c^{-\epsilon}\left[P_{qq}(z)G_{q/p}(x_1/z)G_{q/p}(x_2)
\right.\nonumber\\&& \left.+
P_{qq}(z)G_{q/p}(x_1)G_{q/p}(x_2/z)\right]
\frac{dz}{z}\left(\frac{1-z}{z}\right)^{-\epsilon}dx_1dx_2,
\end{eqnarray}
where $G_{q/p}(x)$ is the bare parton distribution function (PDF).

\subsubsection{Massless antiquark emission}

In addition to the real gluon emission, a second set of real emission
corrections to the inclusive cross section for $pp\rightarrow tt$
 at NLO involves the processes with an additional massless antiquark
$\bar u$ in the final state. Since the contributions from real massless $\bar u$ emission
contain initial state collinear singularities, we need to use
the two cutoff phase space slicing method \cite{Harris:2001sx} to
isolate these collinear divergences. The
cross sections for the processes with an additional massless
$\bar{u}$ in the final state can be expressed as
\begin{eqnarray}
\label{sigma:nc} d\sigma^{add} & = &\Big\{d\hat{\sigma}^{\overline{C}}
(u g \rightarrow tt +
\bar{u})G_{u/p}(x_1)G_{g/p}(x_2)+ \nonumber\\ && d\hat{\sigma}^B
\left[\frac{\alpha_s}{2\pi}\frac{\Gamma(1-\epsilon)}{\Gamma(1-2\epsilon)}
\left(\frac{4\pi\mu^2_r}{s}\right)^{\epsilon}\right]
(-\frac{1}{\epsilon}) \delta_c^{-\epsilon}
P_{qg}(z)G_{u/p}(x_1/z)G_{g/p}(x_2) \nonumber\\
&& \frac{dz}{z}\left(\frac{1-z}{z}\right)^{-\epsilon}+ (x_1
\leftrightarrow x_2)\Big\}dx_1dx_2,
\end{eqnarray}
where
\begin{eqnarray}
P_{qg}(z) =
\frac{1}{2}[z^2+(1-z)^2].
\end{eqnarray}
The $\hat{\sigma}^{\overline{C}}$ terms in Eq. (\ref{sigma:nc})
represents the noncollinear cross sections for the $qg$
initiated processes which can be written in the form
\begin{eqnarray}
d\hat{\sigma}^{\overline{C}}=\frac{1}{2s}\Big\{
|\mathcal{M}(ug \stackrel{\text{noncollinear}}\longrightarrow tt+\bar{u})|^2\Big\}d{\overline{\Gamma}}_3,
\end{eqnarray}
where $d\overline{\Gamma}_3$ is the three-body phase space in the
noncollinear region. The other terms in Eq. (\ref{sigma:nc}) are the
collinear singular cross sections.

\subsubsection{Mass factorization}
The soft divergences can be canceled out after adding the
 renormalized virtual corrections and the real
corrections together. However, there still remain some collinear
divergences which should be absorbed into a redefinition of the PDFs at the
NLO \cite{Altarelli:1979ub,Collins:1989gx}. This procedure means that first we
convolute the partonic cross section with the bare PDF
$G_{\alpha/p}(x)$ and then use the renormalized PDF
$G_{\alpha/p}(x,\mu_f)$ to replace $G_{\alpha/p}(x)$. In the modified minimal subtraction
($\overline{\text{MS}}$) convention the scale-dependent PDF
$G_{\alpha/p}(x,\mu_f)$ is given by \cite{Harris:2001sx}
\begin{eqnarray}
\label{modifiedPDF} G_{\alpha/p}(x,\mu_f) & = & G_{\alpha/p}(x) +
\sum_{\beta}\left(-\frac{1}{\epsilon}\right)\left[
\frac{\alpha_s}{2\pi}\frac{\Gamma(1-\epsilon)}{\Gamma(1-2\epsilon)}
\times \left(\frac{4\pi\mu^2_r}{\mu_f^2}\right)^{\epsilon}\right]\nonumber\\
&& \times \int_x^1 \frac{dz}{z} P_{\alpha\beta}(z) G_{\beta/p}(x/z).
\end{eqnarray}
Then the $\mathcal{O}
(\alpha_s)$ expression for the remaining collinear contribution
can be written in the following form
\begin{eqnarray}
&& d\sigma^{coll}=  d\hat{\sigma}^B\bigg[\frac{\alpha_s}{2\pi}
\frac{\Gamma(1-\epsilon)} {\Gamma(1-2\epsilon)}
\bigg(\frac{4\pi\mu^2_r}{s}\bigg)^\epsilon \bigg] \left\{ \right.
\tilde{G}_{u/p}(x_1,\mu_f) G_{u/p}(x_2,\mu_f)+
\\ && \hspace{1.2cm}
G_{u/p}(x_1,\mu_f) \tilde{G}_{u/p}(x_2,\mu_f)
+2 \big(\frac{A_1^{sc}(u\rightarrow
u g)}{\epsilon} +A_0^{sc}(u\rightarrow u
g)\big) \nonumber
\\ && \hspace{1.2cm}
G_{u/p}(x_1,\mu_f) G_{u/p}(x_2,\mu_f)
\left. \right\} dx_1dx_2,\label{11}
\end{eqnarray}
where
\begin{eqnarray}
A_0^{sc}&=&A_1^{sc}\ln(\frac{s}{\mu_f^2}), \\
A_1^{sc}(q\rightarrow qg)&=&C_F(2\ln\delta_s +3/2), \\
\tilde{G}_{\alpha/p}(x,\mu_f)&=&\sum_{\beta}\int_x^{1-
\delta_s\delta_{\alpha\beta}} \frac{dy}{y}
G_{\beta/p}(x/y,\mu_f)\tilde{P}_{\alpha\beta}(y),
\end{eqnarray}
with
\begin{eqnarray}
\tilde{P}_{\alpha\beta}(y)=P_{\alpha\beta}(y) \ln(\delta_c
\frac{1-y}{y} \frac{s}{\mu_f^2}).
\end{eqnarray}
Then, the IR divergences of the real corrections can be written as
\begin{eqnarray}\label{eq2}
 \hat\sigma^{R(IR)}_{pro}&=& \frac{\alpha_s C_\epsilon C_F}{\pi}\left\{    \right.
(\frac{1}{\epsilon^2}+(\frac{5}{2}-2 a_4)\frac{1}{\epsilon})\hat \sigma_t)+
(\frac{1}{\epsilon^2}+(\frac{5}{2}-2 a_3)\frac{1}{\epsilon})\hat \sigma_u)-
\nonumber \\ &&
(\frac{1}{\epsilon^2}+(a_1+\frac{a_2 \beta }{2 \sqrt{1-\beta }}-
\frac{a_2}{\sqrt{1-\beta }}-2 a_3-2 a_4+\frac{5}{2})\frac{1}{\epsilon})\hat \sigma_{tu})
 \left. \right\},
\end{eqnarray}
and now all the IR divergences from the virtual corrections
in Eq.~(\ref{eq1}) are canceled by those in Eq.~(\ref{eq2}) exactly.

\subsection{Real corrections for top quark decay}
The real corrections for top quark decay only consist of the radiations of an
additional gluon $t \longrightarrow b\ W^+\ g  \longrightarrow b\ e^+\ \nu\ g$.
Following the same procedure as for the real gluon emission of the
production process, we can write down the soft and collinear parts easily
\begin{eqnarray}\label{eq3}
d\hat\sigma_{soft}&=&\frac{C_F \alpha _s}{2 \pi }C_\epsilon \{(\frac{1}{\epsilon^2}+
\frac{(1-2\ln(\delta_s))}{\epsilon})+2(\ln(\delta_s)^2-
\ln(\delta_s)+2)-\frac{\pi^2}{6}\}d\hat\sigma_{B}
\\ \nonumber
d\hat\sigma_{coll}&=&\frac{C_F \alpha _s}{2\pi} C_\epsilon
\{(2 \ln \left(\frac{m_t^2}{2 {p_b . p_t}}\right)+2 \ln (\delta_s)+\frac{3}{2})\frac{1}{
\epsilon}-2 \ln \left(\delta _c\right) \ln \left(\frac{m_t^2}{2 p_b . p_t}\right)
\\ \nonumber &&
-2 \ln \left(\delta _c\right) \ln \left(\delta _s\right)-\frac{3 \ln
 \left(\delta _c\right)}{2}-2 \ln \left(\delta _s\right)
 \ln \left(\frac{m_t^2}{2 p_b . p_t}\right)-\ln ^2\left(\frac{m_t^2}{2 p_b . p_t}\right)
\\ \nonumber &&
 -\ln ^2\left(\delta _s\right)-\frac{\pi ^2}{3}+3
\}d\hat\sigma_{B}.
\end{eqnarray}
And then the total IR divergent parts are
\begin{eqnarray}\label{eqdecr}
 \Gamma^{R(IR)}_{dec}&=& \frac{\alpha_s C_\epsilon C_F}{2 \pi}\
(\frac{1}{\epsilon^2}+(5+4 \ln(\frac{m_{t}^2}{2 p_b . p_t}))\frac{1}{2\epsilon})\Gamma_t,
\end{eqnarray}
which cancel the IR divergence in Eq.~(\ref{eqdecv}).

\section{Numerical Results}\label{s5}
In this section we present all the numerical results. We fix the top mass
$m_t=172.5 ~{\rm GeV}$ and all
 the other SM input parameters are taken
to be~\cite{Nakamura:2010zzi}:
\begin{equation}
\quad \alpha_s(M_Z)=0.118, \quad \alpha(M_Z)=1/128.921,  \quad M_W=80.399~\text{GeV},
\end{equation}
The CTEQ6.6 PDF set~\cite{Pumplin:2002vw} is used throughout the calculations except for the backgrounds in which the CTEQ6L~\cite{Pumplin:2002vw} is used. Both the renormalization and factorization
scales are fixed to the top quark mass. We adopt the same cuts as in Ref.~\cite{Chatrchyan:2011dk}
\begin{eqnarray}
&&p_T(j)>30~{\rm GeV},\quad |\eta (j)|<2.5,\nonumber\\
&&|\eta (l)|<2.4,\quad
\quad \not{\! E}_T^{e\mu(\text{ee},  \mu \mu)}>30(20)~{\rm GeV},
\end{eqnarray}
and the same-sign isolated leptons have $p_T(l)>10~{\rm GeV}$, one of which should have $ p_T(l)>20~{\rm GeV}$.

\subsection{QCD NLO Results}

\begin{figure}[H]
  \subfigure{
    \begin{minipage}[b]{0.5\textwidth}
      \begin{center}
     \scalebox{0.3}{\includegraphics*{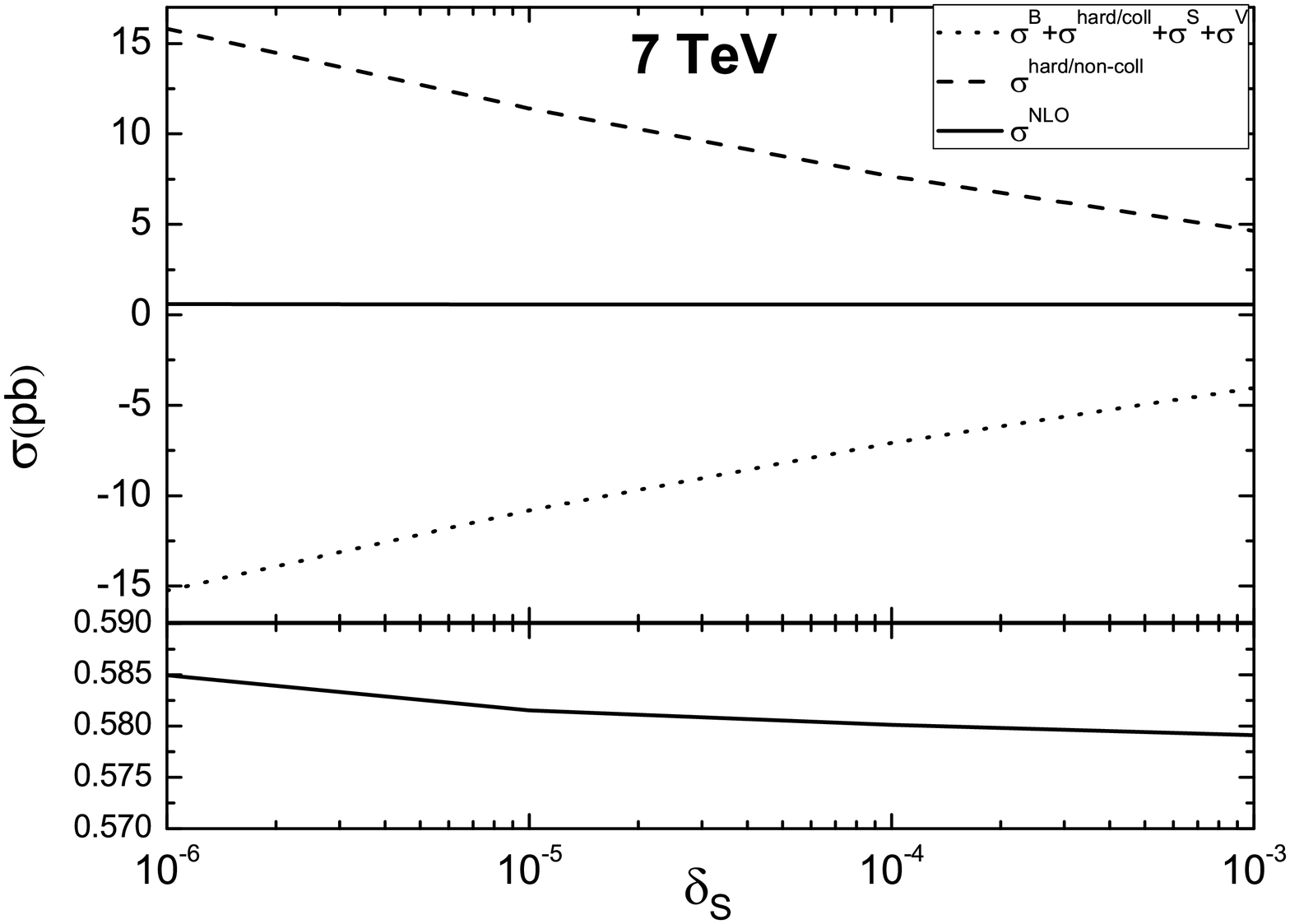}}
      \end{center}
    \end{minipage}}
  \subfigure{
    \begin{minipage}[b]{0.5\textwidth}
      \begin{center}
     \scalebox{0.3}{\includegraphics*{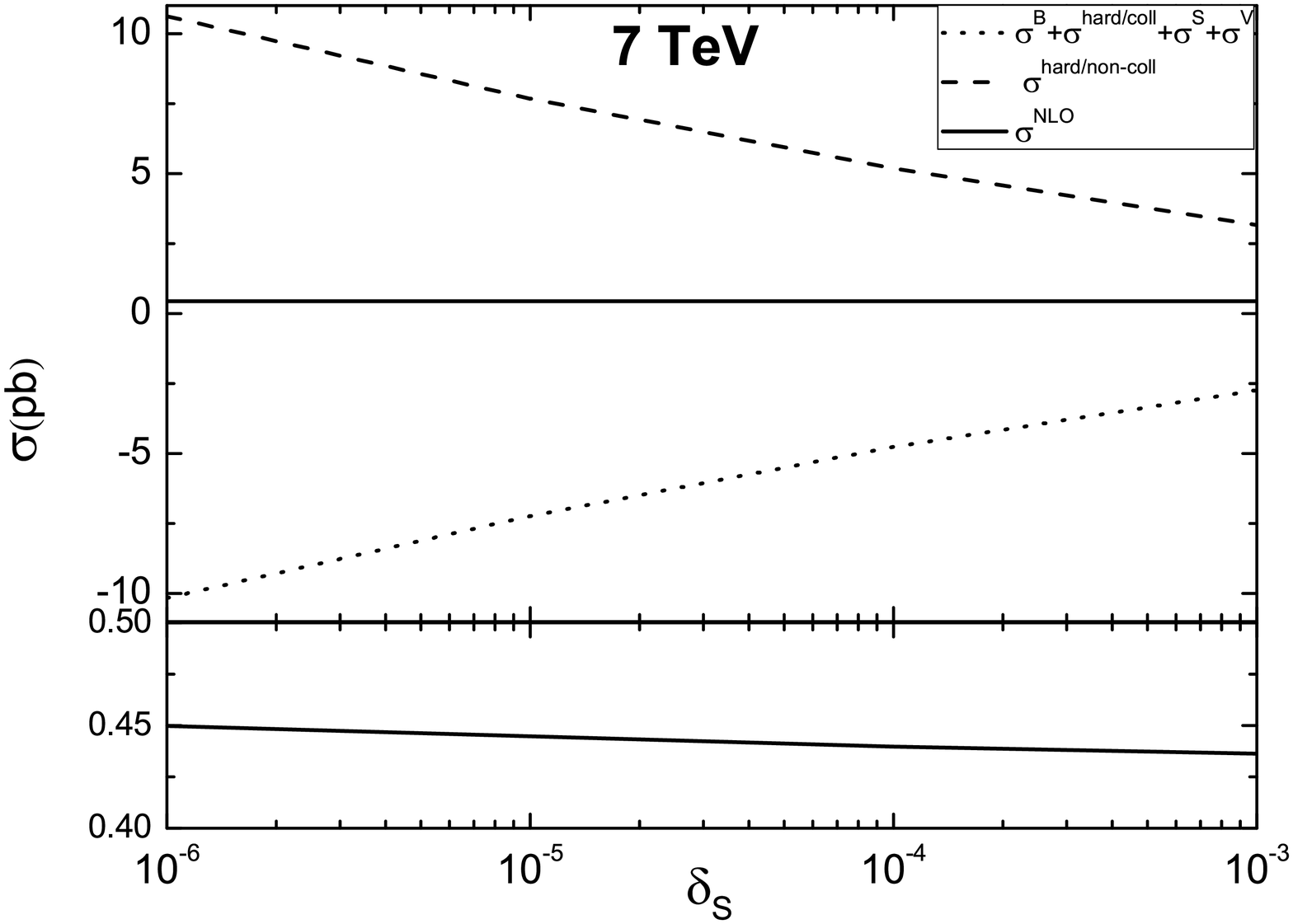}}
      \end{center}
    \end{minipage}}
    \caption{\label{deltas} Inclusive total cross sections for $pp\rightarrow tt \rightarrow e^{+}e^{+}+2jets
    +\slashed E+X$ at the
     LHC as a function of $\delta_{s}$ in the phase space slicing treatment. The left and right figures
      are shown for the cases without and with cuts, respectively. Here, $C_R=2$ and the $\delta_{c}$  is chosen to be $\delta_{c}=\delta_{s}/50$.}
\end{figure}

We can use the NLO corrections to update the LO results, but
before that, we would show that it is reasonable to use the two cutoff
phase space slicing method in our calculation. To use the two cutoff method,
we introduce two small cutoffs $\delta_s$ and $\delta_c$. For the total cross section without any kinematic cuts imposed,
the $\delta_s$ dependence is shown in Fig.\ref{deltas} (left).
If we impose the cuts on the parton, it would not be infrared safe.
So we adopt the anti-$k_t$ jet algorithm~\cite{Cacciari:2008gp} and set $R=0.7$ in our calculation.
After all cuts imposed, the $\delta_s$ dependence is shown in Fig.\ref{deltas} (right).
Here, we only show the electron final state as an example. The soft,
collinear and the noncollinear contributions individually depend
strongly on the cutoffs. However, the cutoff dependence in the two
contributions ($\sigma^{S} + \sigma^{coll}$ and
$\sigma^{\overline{HC}} + \sigma^{\overline{C}}$) nearly cancel
each other, so the final results for $\sigma^{NLO}$ are almost
independent of the cutoffs.

\begin{figure}[h]
      \begin{center}
     \scalebox{0.5}{\includegraphics*{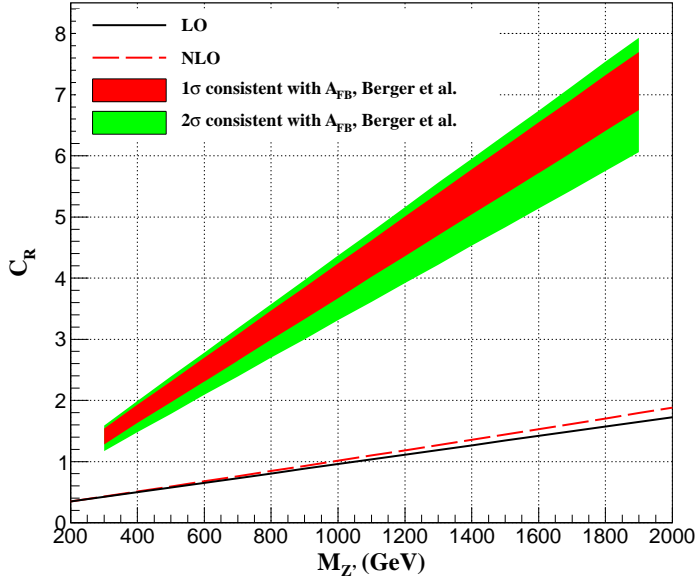}}
      \end{center}
  \caption[]{\label{cexp} Constraint line~(solid) from the CMS collaboration. Considering the NLO corrections,
  we have a new constraint line~(dashed), which loosen the constraint on the model parameters.}
\end{figure}
As mentioned in the introduction, the CMS Collaboration has set a limit on the parameter
$C_R$ and $M_{Z^\prime}$ based on
the LO calculations. According to our numerical result,
the NLO corrections can loosen this constraint
but this kind of $Z^\prime$ still cannot explain the top quark forward-backward asymmetry as shown in Fig.~\ref{cexp}.

\begin{figure}
  \subfigure{
    \begin{minipage}[b]{0.3\textwidth}
          \begin{center}
     \scalebox{0.2}{\includegraphics*{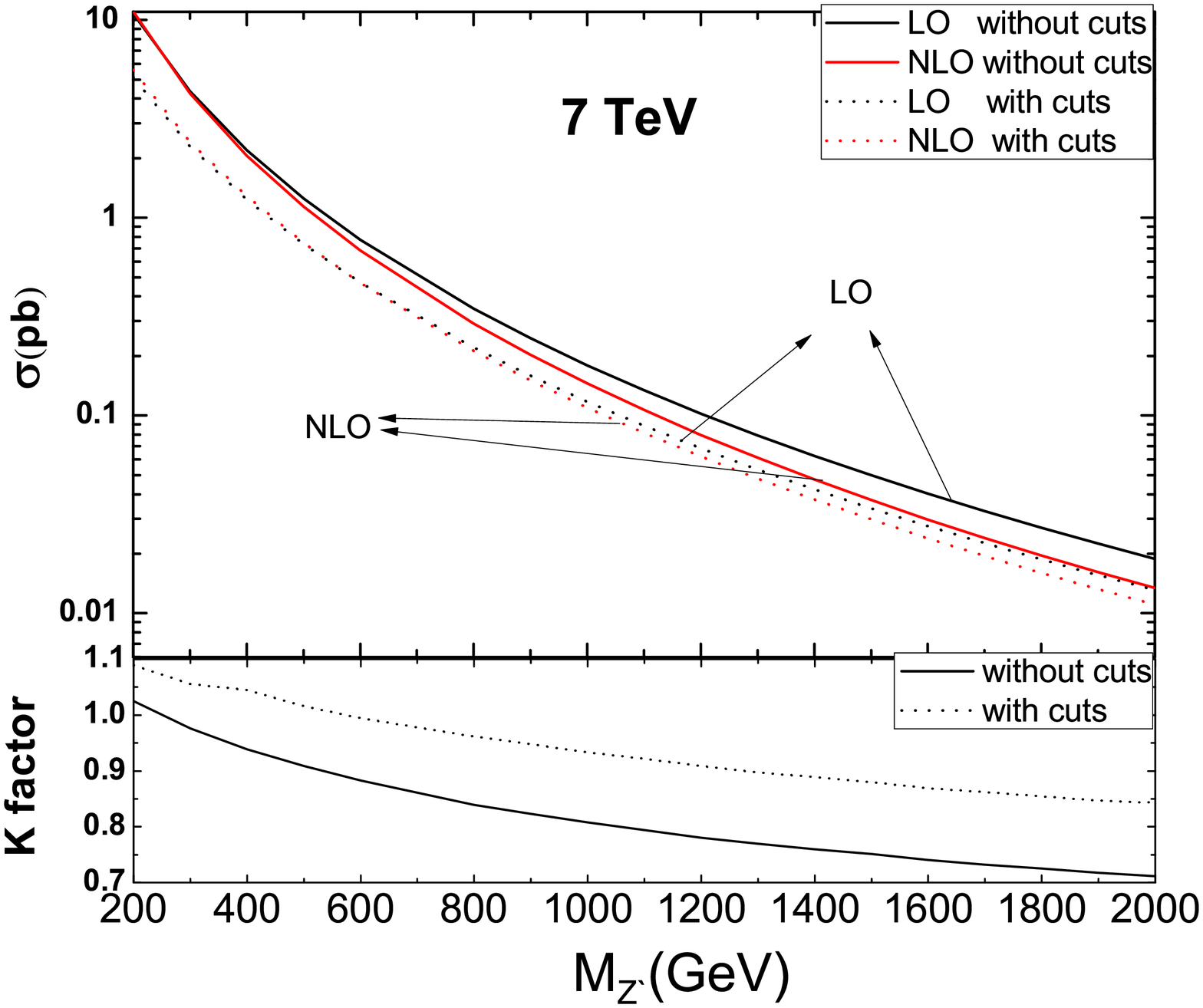}}
           \end{center}
    \end{minipage}}
  \subfigure{
    \begin{minipage}[b]{0.3\textwidth}
          \begin{center}
     \scalebox{0.2}{\includegraphics*{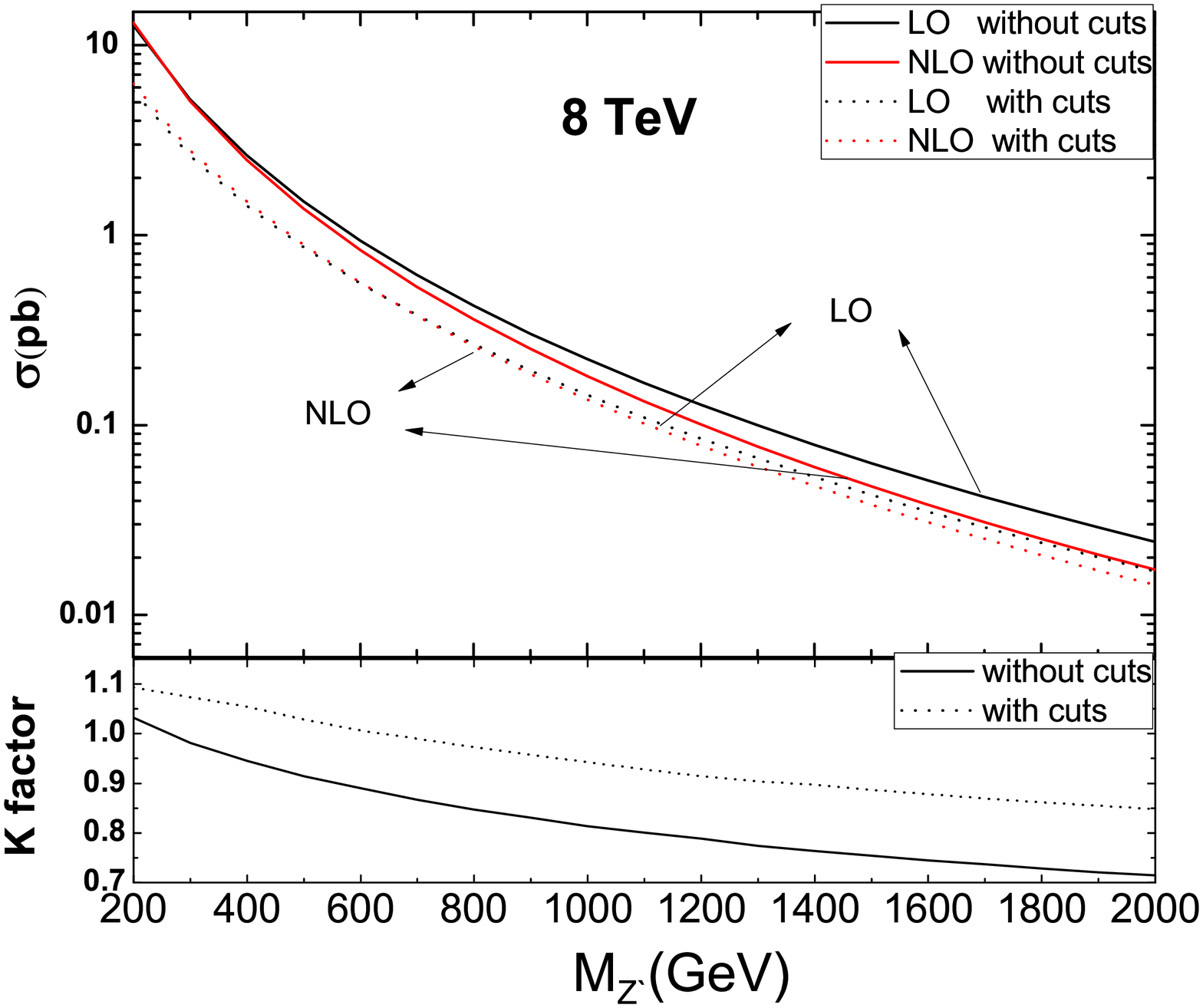}}
           \end{center}
    \end{minipage}}
      \subfigure{
    \begin{minipage}[b]{0.3\textwidth}
          \begin{center}
     \scalebox{0.2}{\includegraphics*{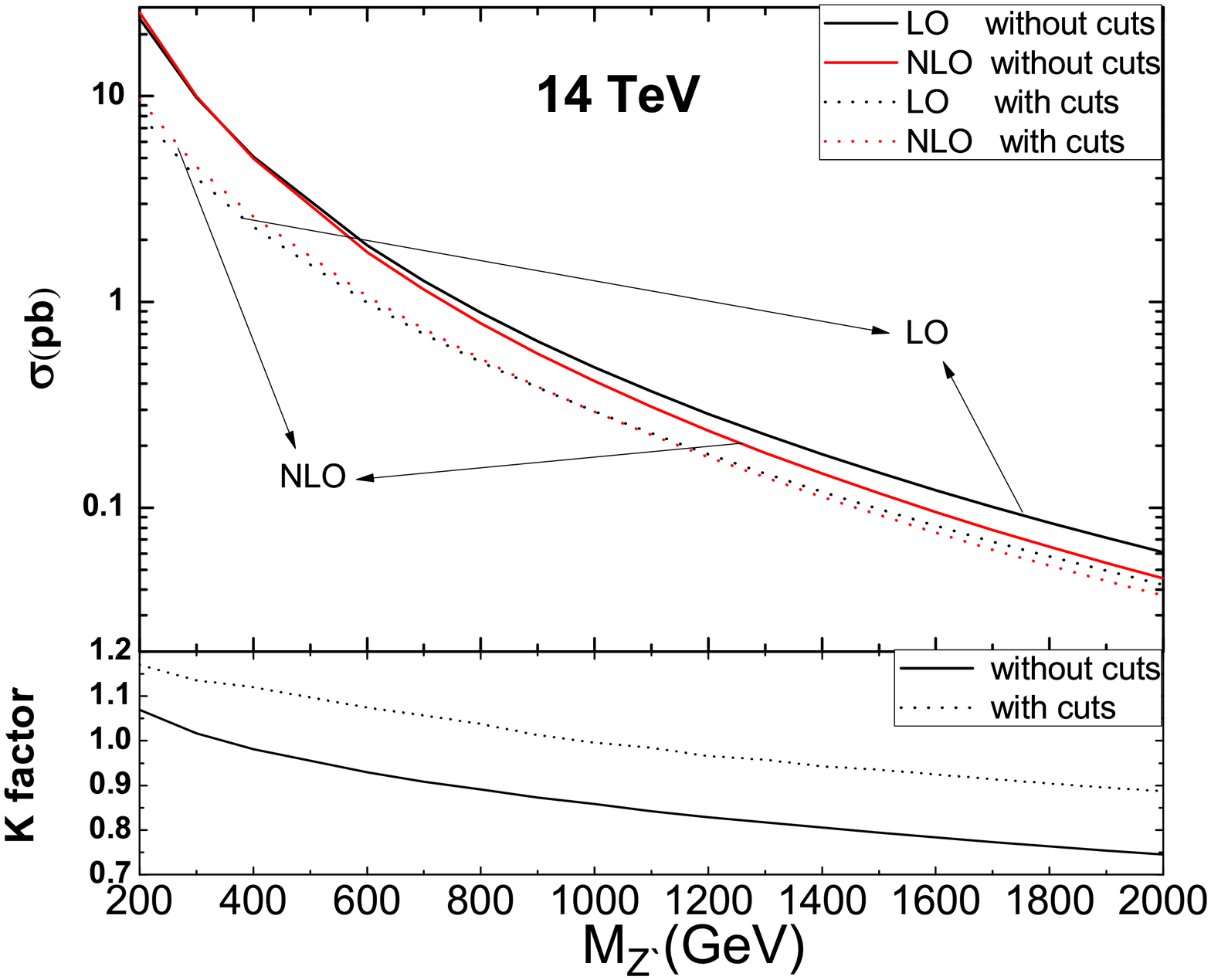}}
          \end{center}
    \end{minipage}}
    \caption{\label{zmassrun} The total cross section and the K factor as the function of the $M_{Z^\prime}$ at the LHC with $C_R=1$.}
\end{figure}

In Fig.\ref{zmassrun}, we plot the total cross section and the K factor, defined as $\sigma_{NLO}/\sigma_{LO}$, as the function of the $Z^\prime$ boson mass. Because of the large negative contributions from the interference between the loop corrections and the Born amplitudes, the NLO corrections reduce the cross section. We can see that the QCD NLO corrections are more significant for larger $Z^\prime$ boson mass.

\begin{figure}
  \subfigure{
    \begin{minipage}[b]{0.3\textwidth}
      \begin{center}
     \scalebox{0.2}{\includegraphics*{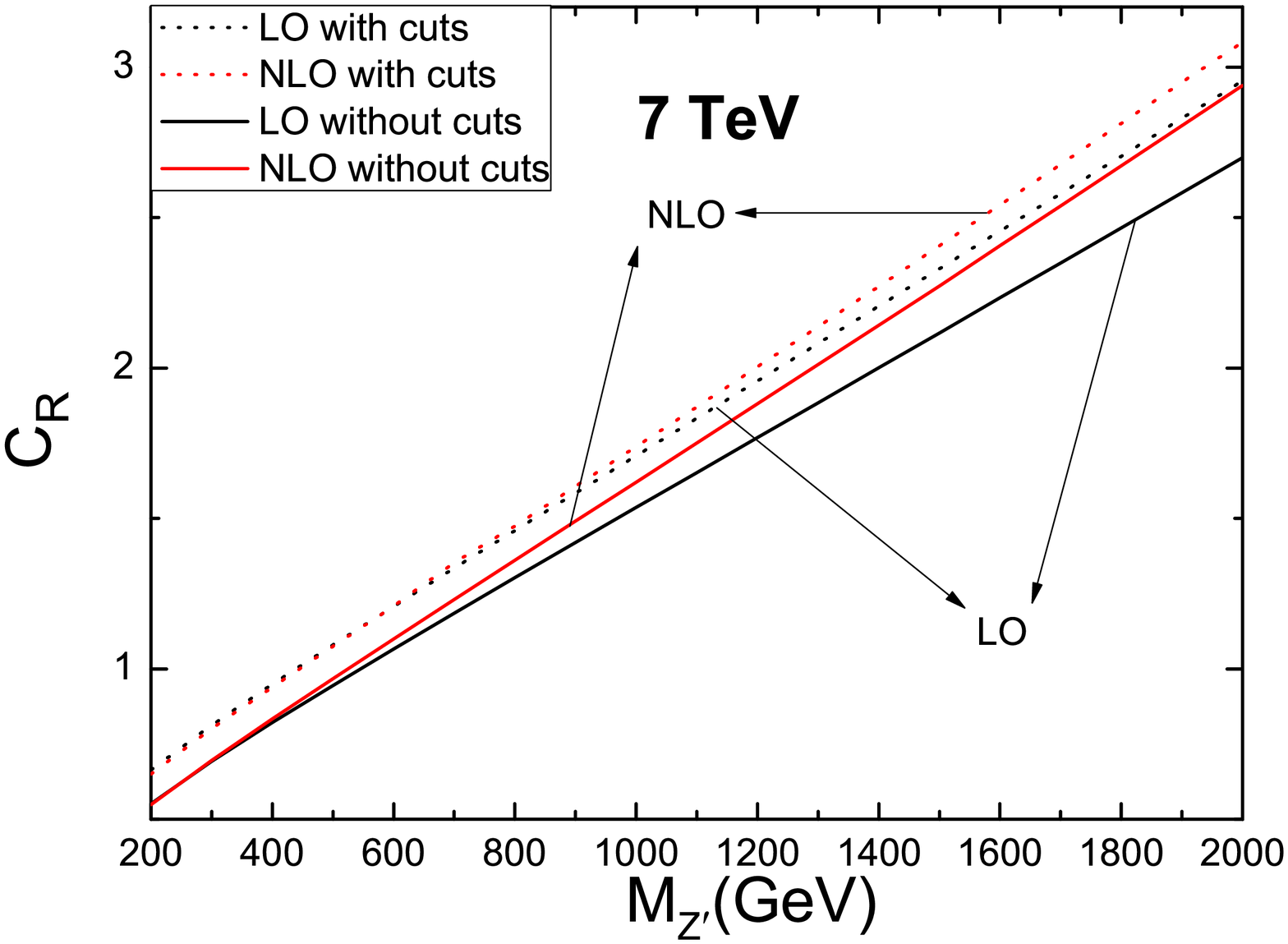}}
      \end{center}
    \end{minipage}}
  \subfigure{
    \begin{minipage}[b]{0.3\textwidth}
      \begin{center}
     \scalebox{0.2}{\includegraphics*{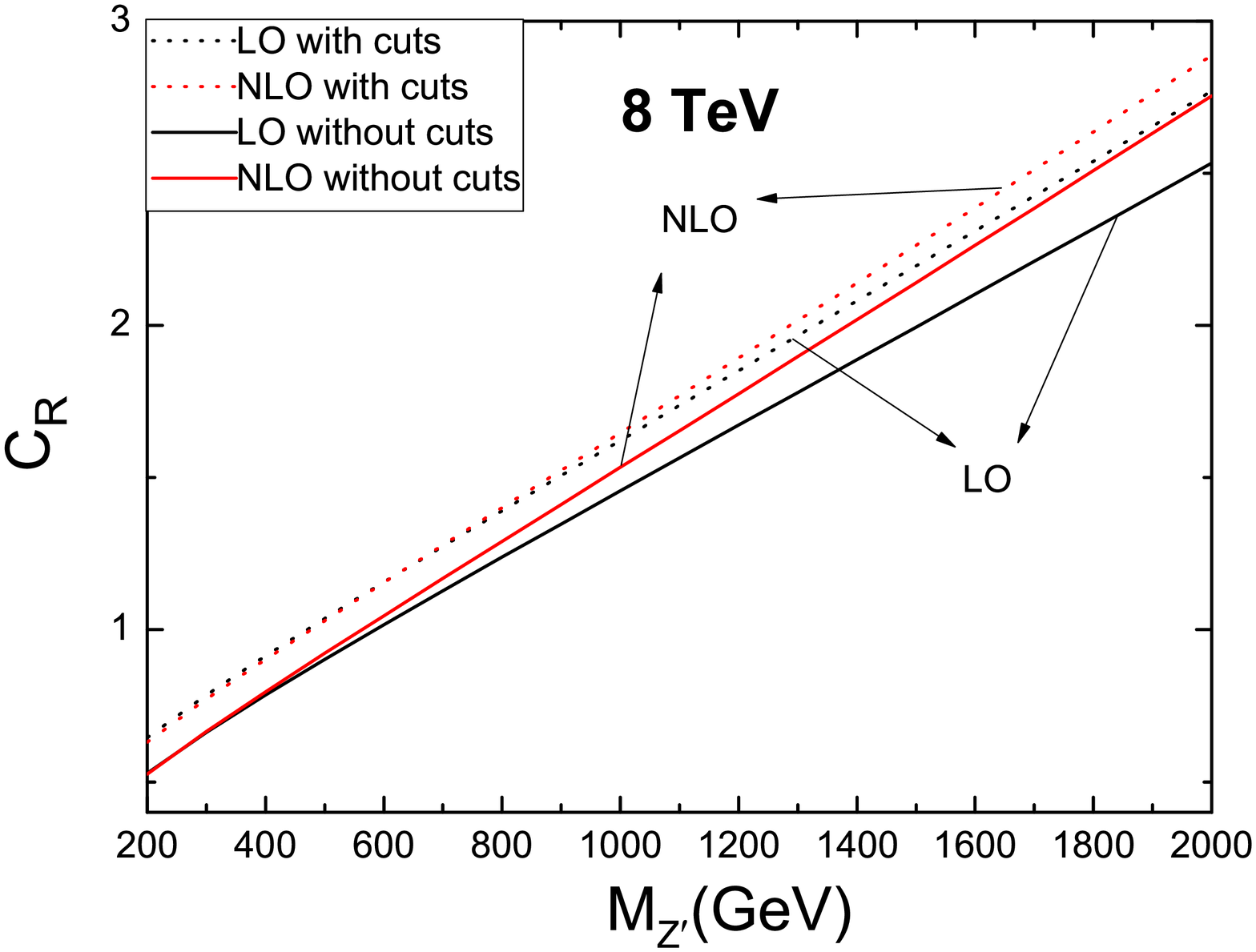}}
      \end{center}
    \end{minipage}}
      \subfigure{
     \begin{minipage}[b]{0.3\textwidth}
      \begin{center}
     \scalebox{0.2}{\includegraphics*{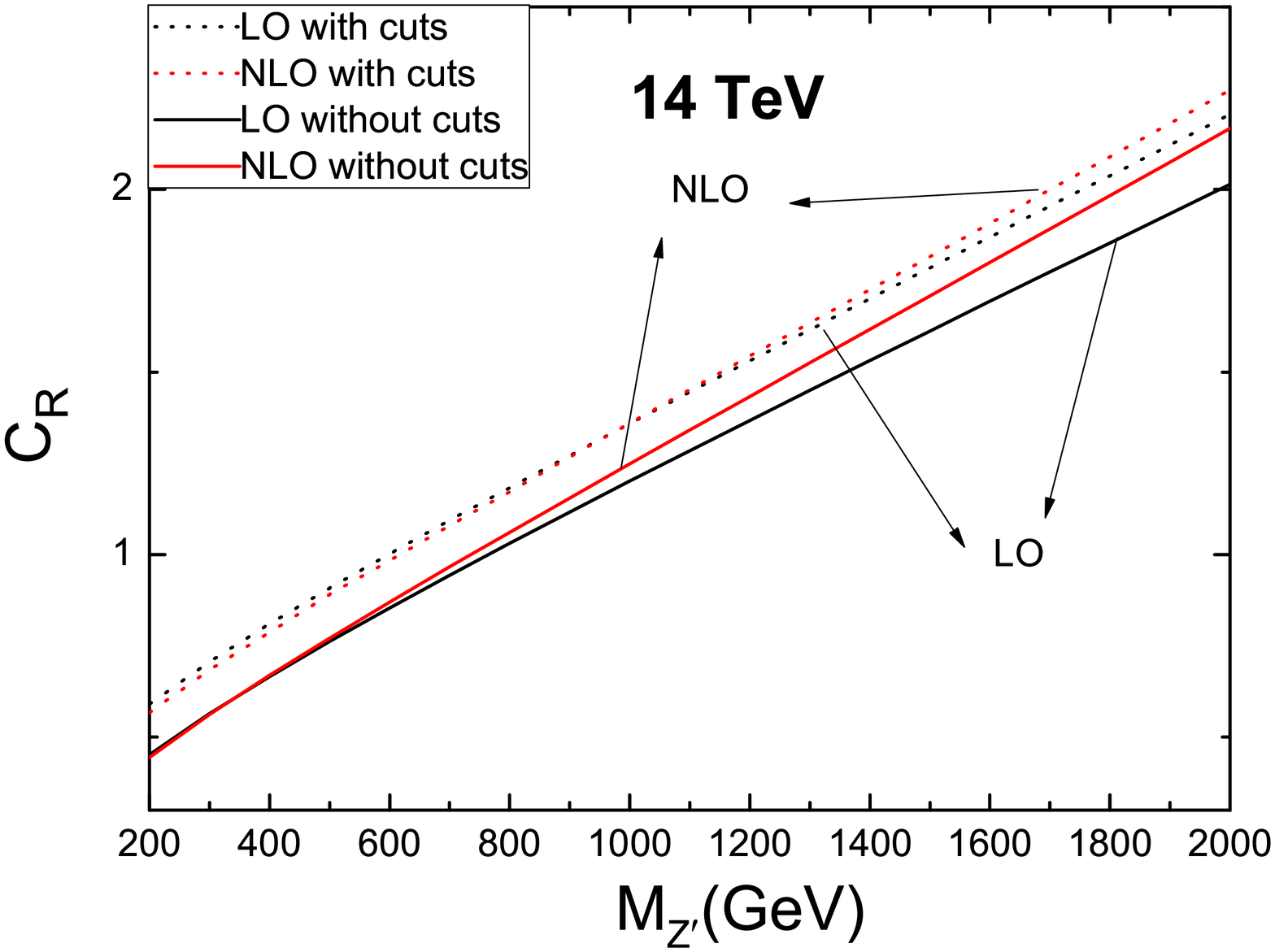}}
      \end{center}
    \end{minipage}}
    \caption{\label{CRzmass} The right-handed coupling $C_R$ versus
    the $Z^{\prime}$ mass. Here we fix the total cross sections to be 1~pb.}
\end{figure}

If the total cross section is fixed to be a certain value, for example 1 pb,
we can also plot the $C_R$ parameter as the function of the $Z^\prime$ boson mass, as shown in Fig.\ref{CRzmass}.
It is easy to find that, the curve in Fig.\ref{CRzmass} is nearly a straight line,
when $Z^\prime$ boson mass is larger than about 500 GeV.
This means the numerical results can be written in the following form if the $Z^{\prime}$ decay width($\Gamma_{Z^\prime}$) is fixed
to a certain value,
\begin{equation}
\sigma=A \frac{C_R^4}{(M_Z^\prime/\text{GeV}+b)^4},
\end{equation}
where $A$ and $b$ are two parameters that depend on the cuts and the center-of-mass Energy~($E_{CM}$).
Then, when $Z^\prime$ boson mass is large enough, i.e., for $E_{CM}=7, 8~\text{TeV}, M_{Z^\prime}>600 ~\text{GeV}$ and for
$E_{CM}=14 ~\text{TeV}, M_{Z^\prime}>800 ~\text{GeV}$.  Since $Z^\prime$ may have other unknown decay channels~\cite{Jung:2009jz}, its total decay width should be larger than the decay width to
$u\bar{t}$ and $\bar{u}t$.
Here we set $\Gamma_{Z^\prime}=\Gamma_{Z^\prime \to u\bar{t}, \bar{u}t}$ with $C_R=2$, then our results can be present as follows:

without cuts:
\begin{eqnarray}\label{eq3}
\sigma_{7 TeV}^{LO}(M_Z^\prime) &=&\frac{5.41975\times 10^{11} \text{pb}~C_R^4}{(M_Z^\prime/\text{GeV}+316.043)^4}, \qquad
\sigma_{7 TeV}^{NLO}(M_Z^\prime) = \frac{3.38221\times 10^{11} \text{pb}~C_R^4}{(M_Z^\prime/\text{GeV}+236.603)^4}  \nonumber \\
\sigma_{8 TeV}^{LO}(M_Z^\prime) &=&\frac{7.2622\times 10^{11} \text{pb}~C_R^4}{(M_Z^\prime/\text{GeV}+341.16)^4}, \qquad
\sigma_{8 TeV}^{NLO}(M_Z^\prime) = \frac{4.52627\times 10^{11} \text{pb}~C_R^4}{(M_Z^\prime/\text{GeV}+257.476)^4}  \nonumber \\
\sigma_{14 TeV}^{LO}(M_Z^\prime) &=&  \frac{2.10819\times 10^{12} \text{pb}~C_R^4}{(M_Z^\prime/\text{GeV}+438.537)^4}, \qquad
\sigma_{14 TeV}^{NLO}(M_Z^\prime) =  \frac{1.35683\times 10^{12} \text{pb}~C_R^4}{(M_Z^\prime/\text{GeV}+342.761)^4}, \nonumber
\end{eqnarray}

with cuts:
\begin{eqnarray}\label{eq4}
\sigma_{7 TeV}^{LO}(M_Z^\prime) &=&\frac{4.13553\times 10^{11} \text{pb}~C_R^4}{(M_Z^\prime/\text{GeV}+368.894)^4}, \qquad
\sigma_{7 TeV}^{NLO}(M_Z^\prime) = \frac{3.13669\times 10^{11} \text{pb}~C_R^4}{(M_Z^\prime/\text{GeV}+303.198)^4},  \nonumber \\
\sigma_{8 TeV}^{LO}(M_Z^\prime) &=& \frac{5.6705\times 10^{11} \text{pb}~C_R^4}{(M_Z^\prime/\text{GeV}+405.522)^4}, \qquad
\sigma_{8 TeV}^{NLO}(M_Z^\prime) =\frac{4.28346\times 10^{11} \text{pb}~C_R^4}{(M_Z^\prime/\text{GeV}+332.548)^4},
  \nonumber \\
\sigma_{14 TeV}^{LO}(M_Z^\prime) &=&  \frac{1.82195\times 10^{12} \text{pb}~C_R^4}{(M_Z^\prime/\text{GeV}+571.211)^4}, \qquad
\sigma_{14 TeV}^{NLO}(M_Z^\prime) =  \frac{1.39929\times 10^{12} \text{pb}~C_R^4}{(M_Z^\prime/\text{GeV}+474.653)^4}.
\end{eqnarray}

\begin{figure}
  \subfigure{
    \begin{minipage}[b]{0.30\textwidth}
      \begin{center}
     \scalebox{0.22}{\includegraphics*{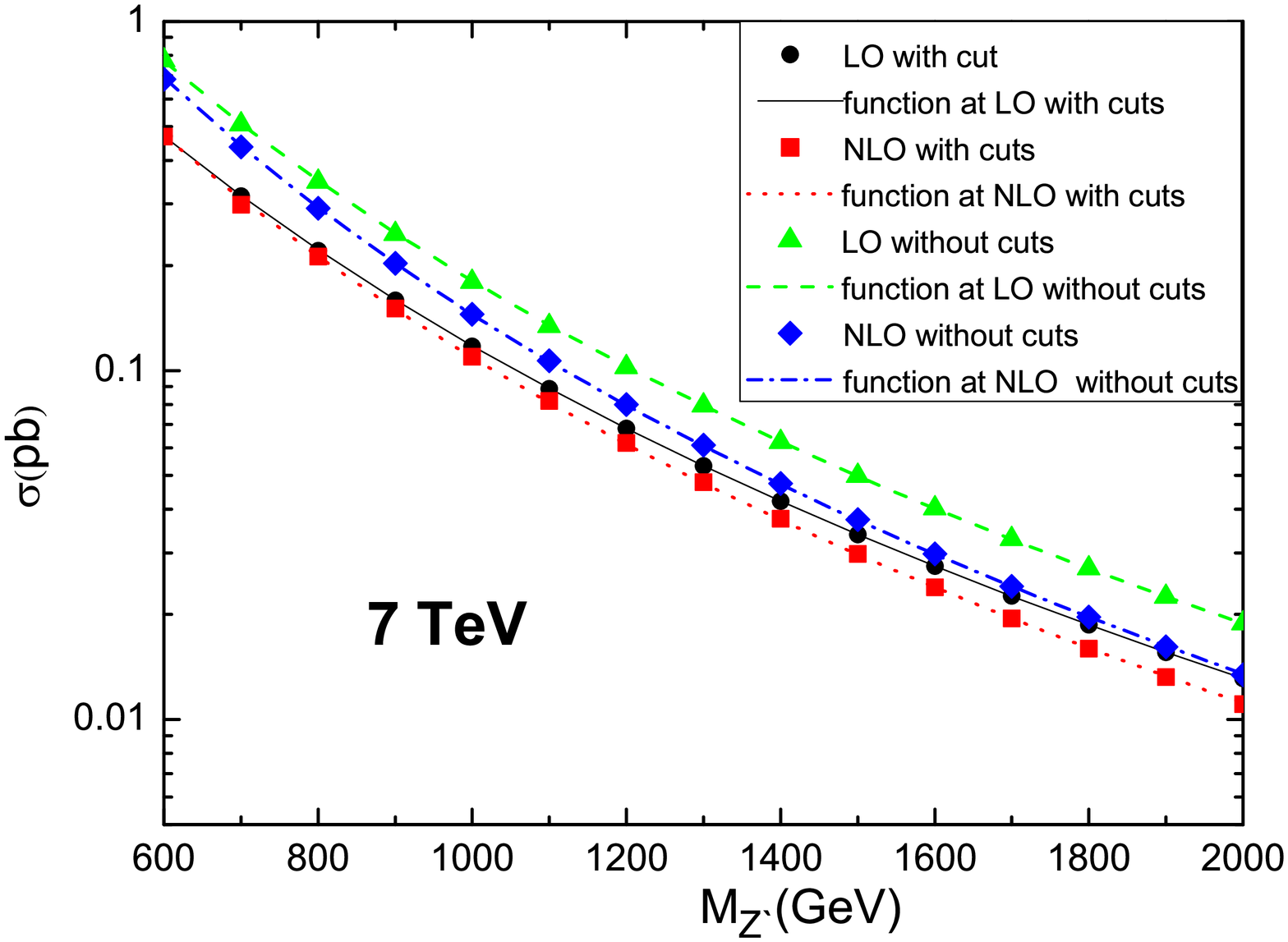}}
      \end{center}
    \end{minipage}}
 \subfigure{
    \begin{minipage}[b]{0.30\textwidth}
      \begin{center}
     \scalebox{0.22}{\includegraphics*{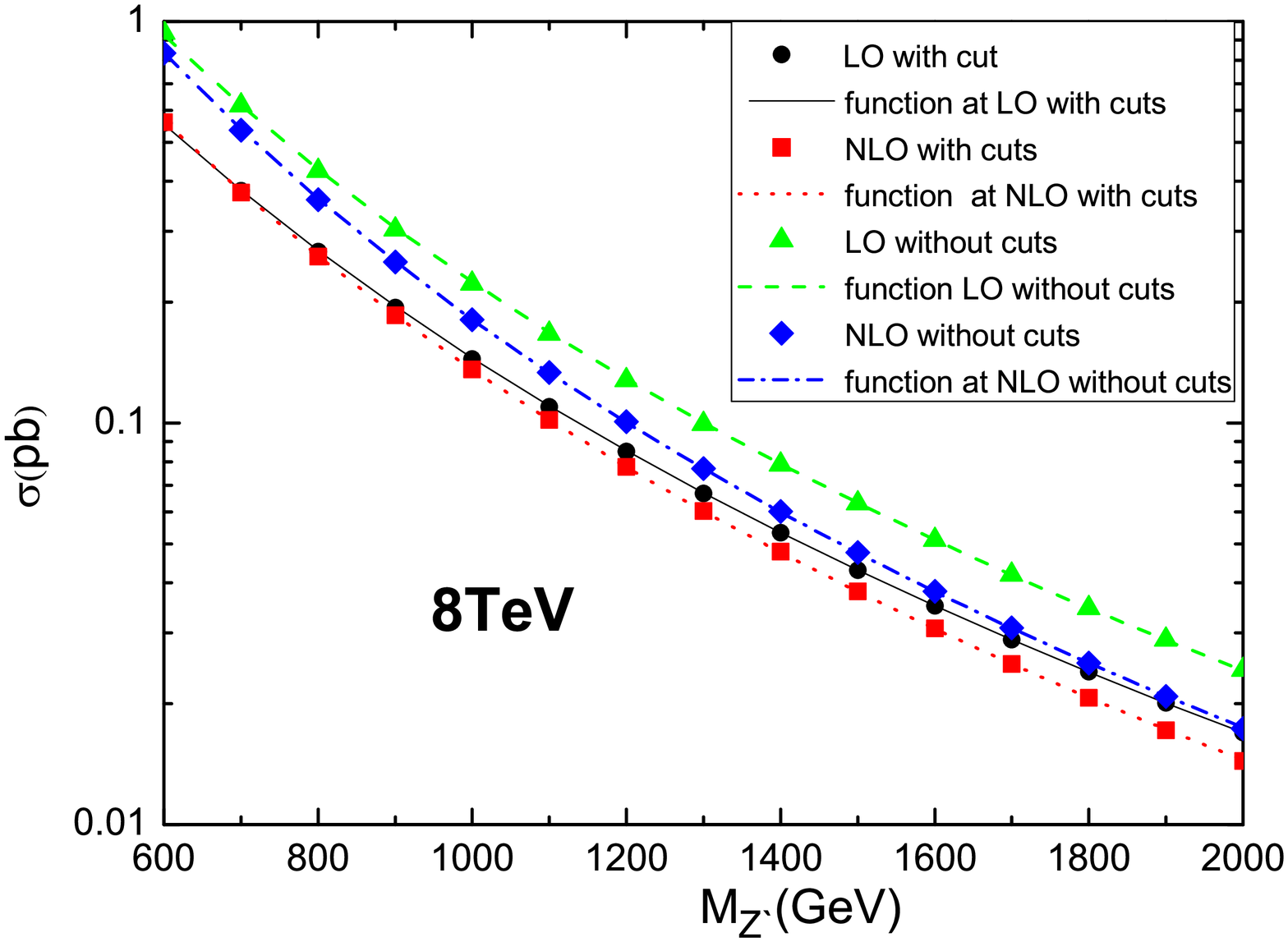}}
      \end{center}
    \end{minipage}}
  \subfigure{
    \begin{minipage}[b]{0.30\textwidth}
      \begin{center}
     \scalebox{0.22}{\includegraphics*{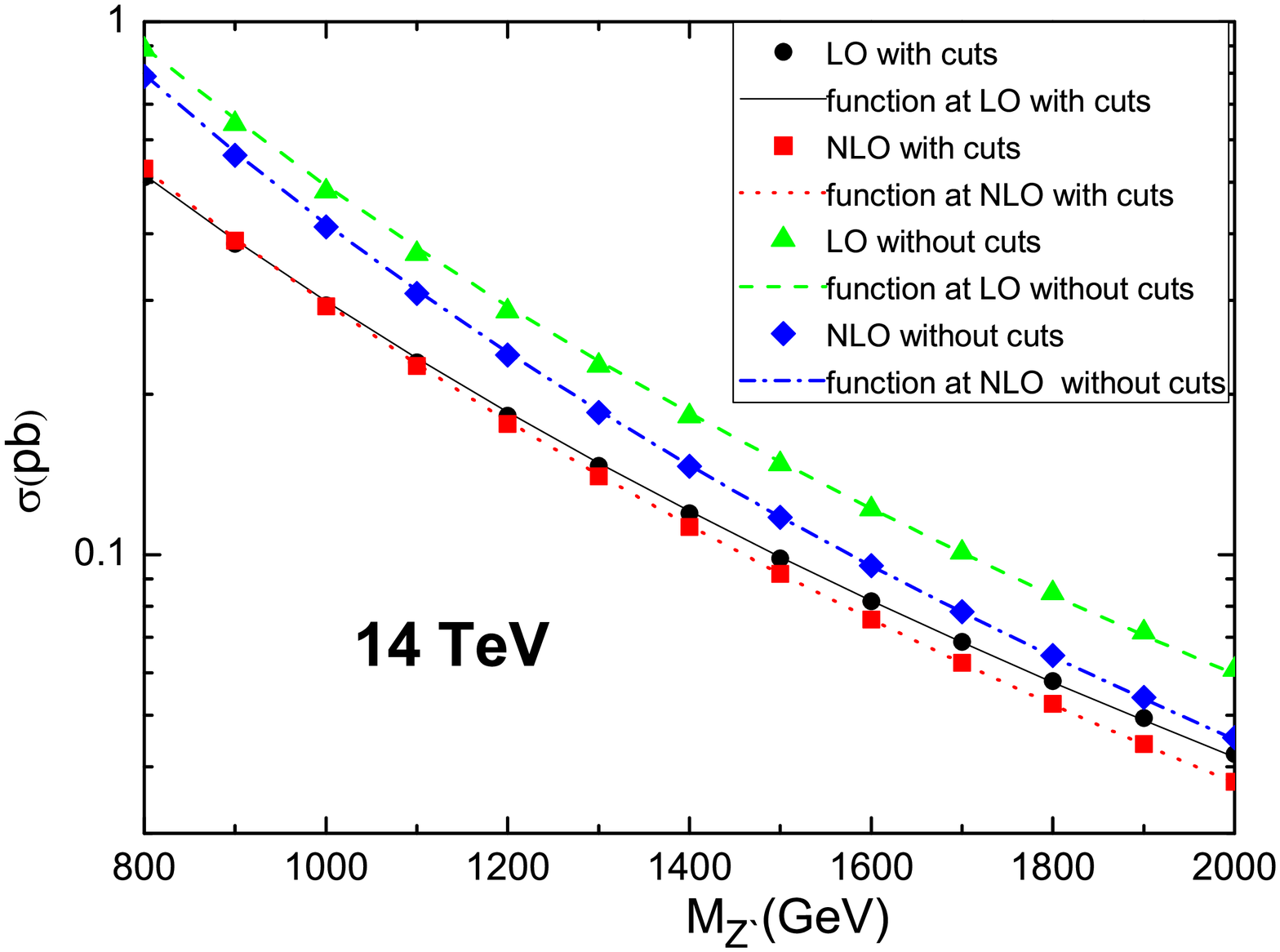}}
      \end{center}
    \end{minipage}}
    \caption{\label{fit} The total cross sections predicted by Eq.(\ref{eq3})~(curve) and the
    numerical results~(scatter points). $C_R$ was chosen to be 1 here.}
\end{figure}

In order to check these functions, we plot them in Fig.~\ref{fit} and find that they are fitted well with
the numerical results of the total cross sections.

\begin{figure}
  \subfigure{
    \begin{minipage}[b]{0.3\textwidth}
     \scalebox{0.201}{\includegraphics*{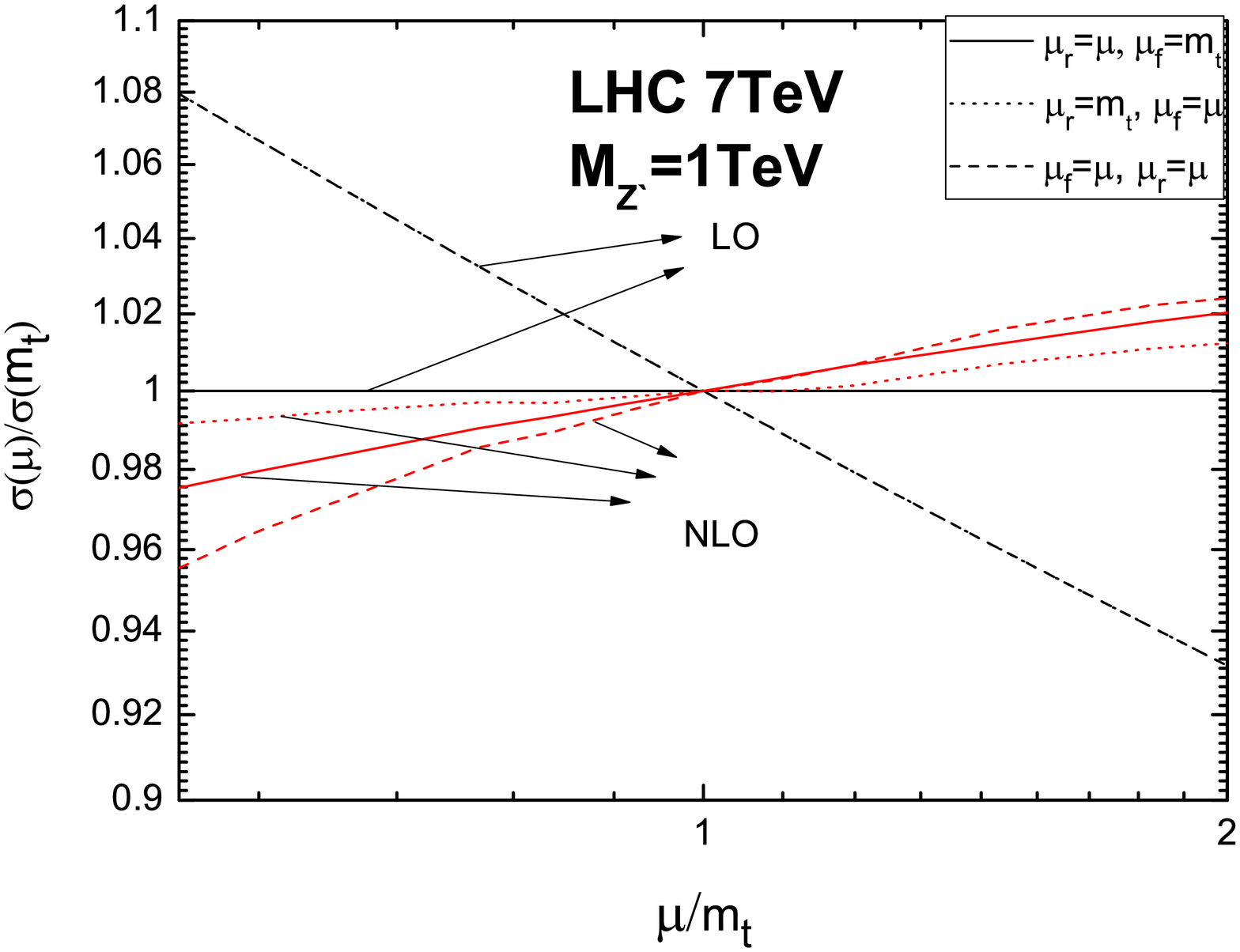}}
    \end{minipage}}
  \subfigure{
    \begin{minipage}[b]{0.3\textwidth}
     \scalebox{0.201}{\includegraphics*{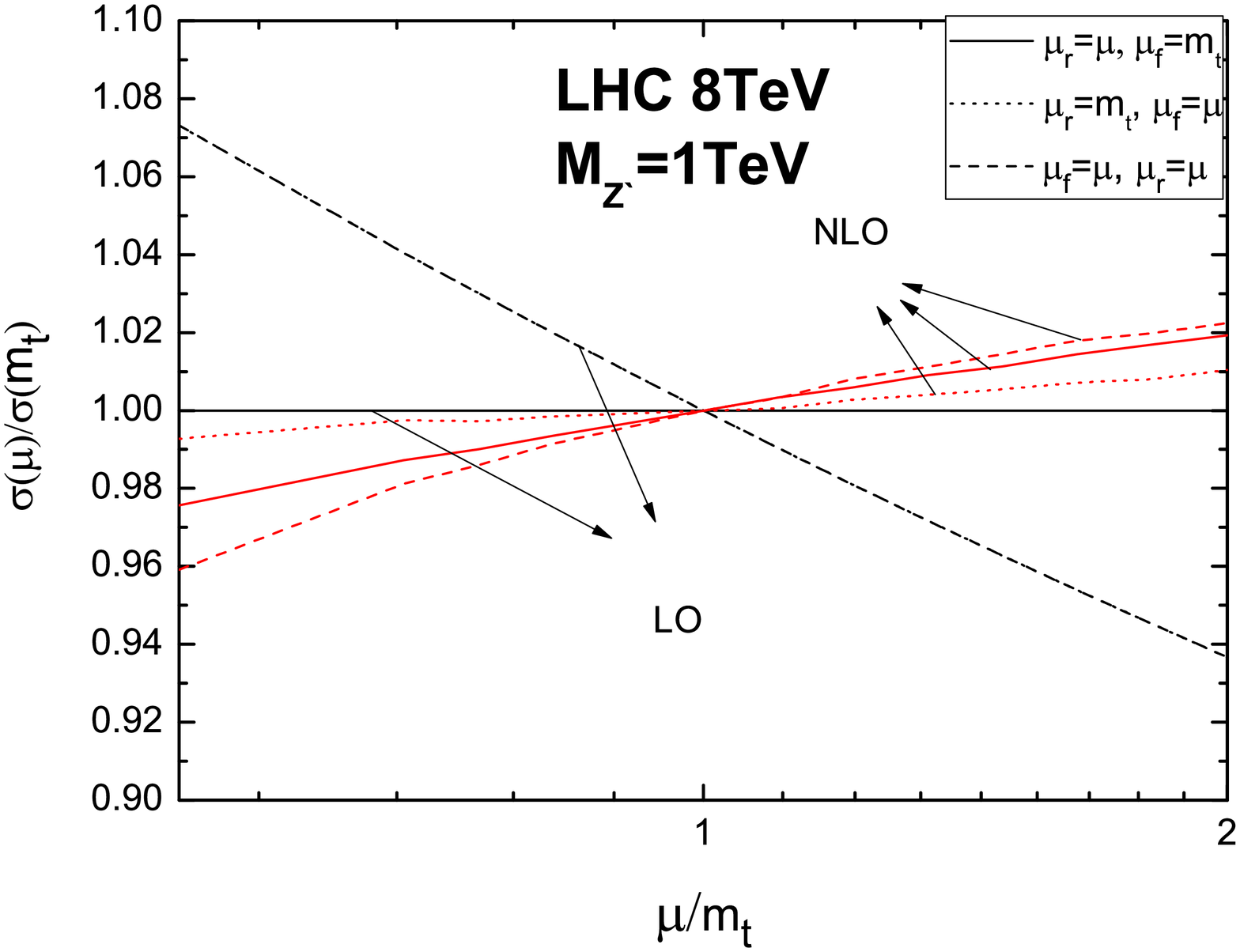}}
    \end{minipage}}
      \subfigure{
    \begin{minipage}[b]{0.3\textwidth}
     \scalebox{0.201}{\includegraphics*{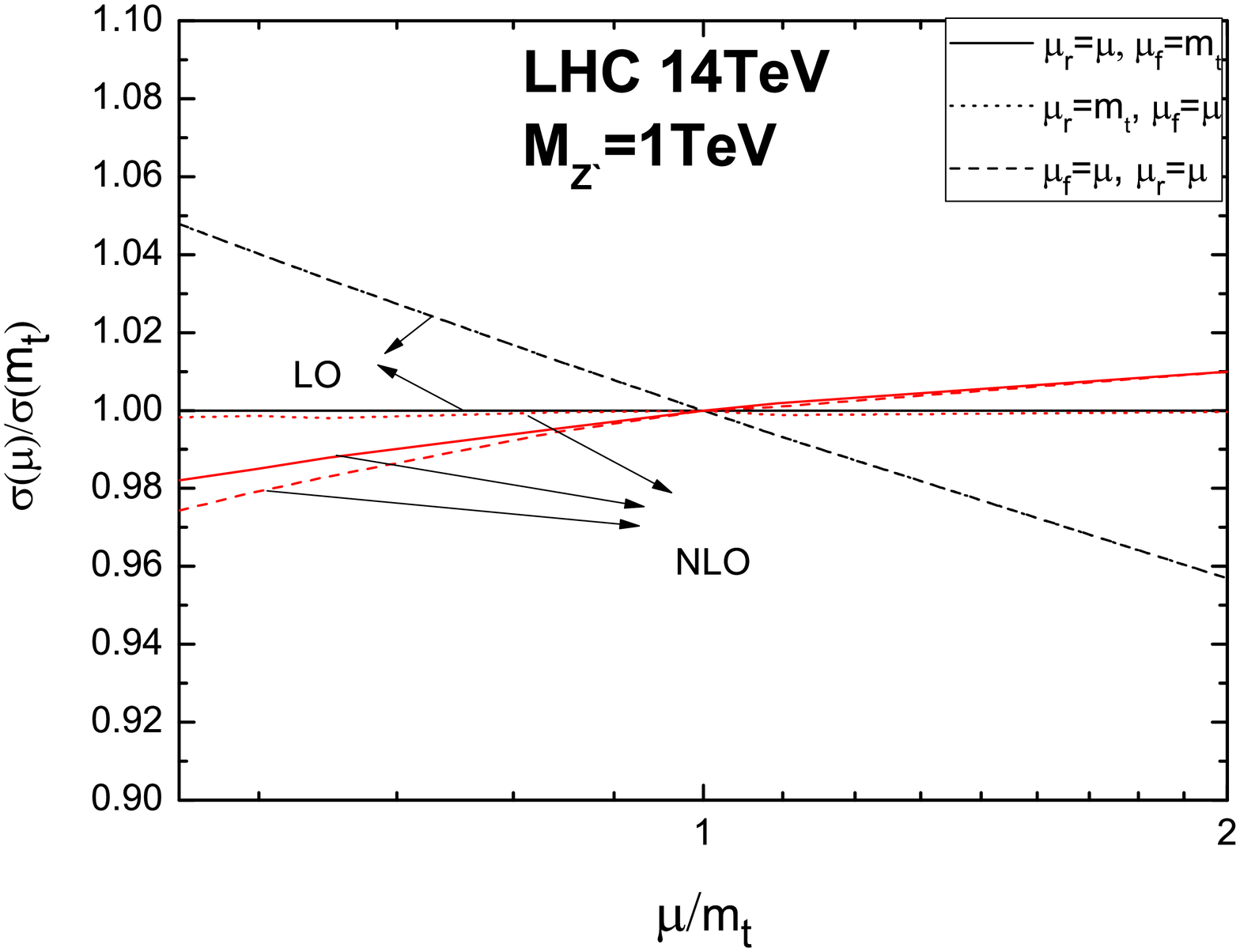}}
    \end{minipage}}
       \subfigure{
    \begin{minipage}[b]{0.3\textwidth}
      \begin{center}
     \scalebox{0.201}{\includegraphics*{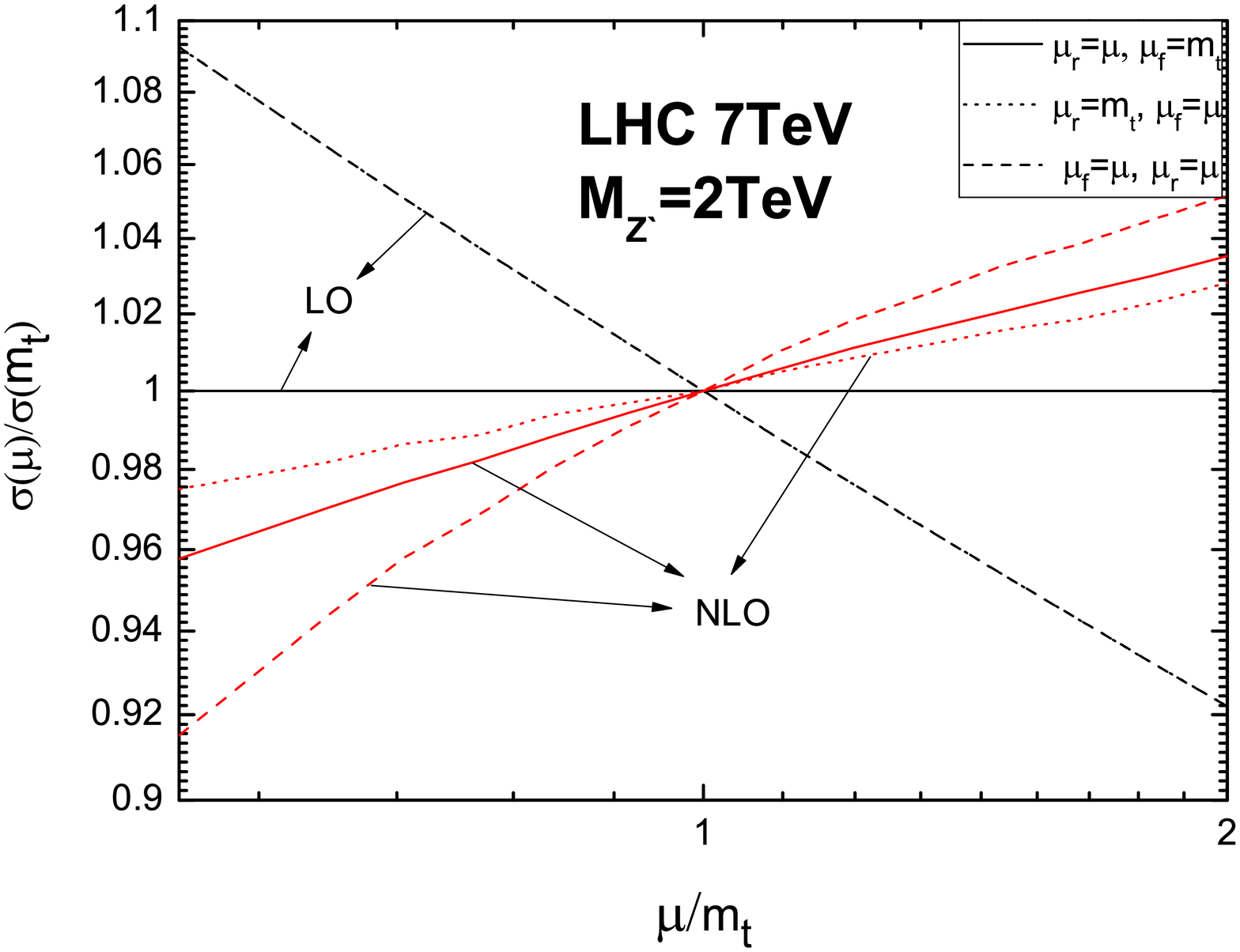}}
      \end{center}
    \end{minipage}}
      \subfigure{
    \begin{minipage}[b]{0.3\textwidth}
      \begin{center}
     \scalebox{0.201}{\includegraphics*{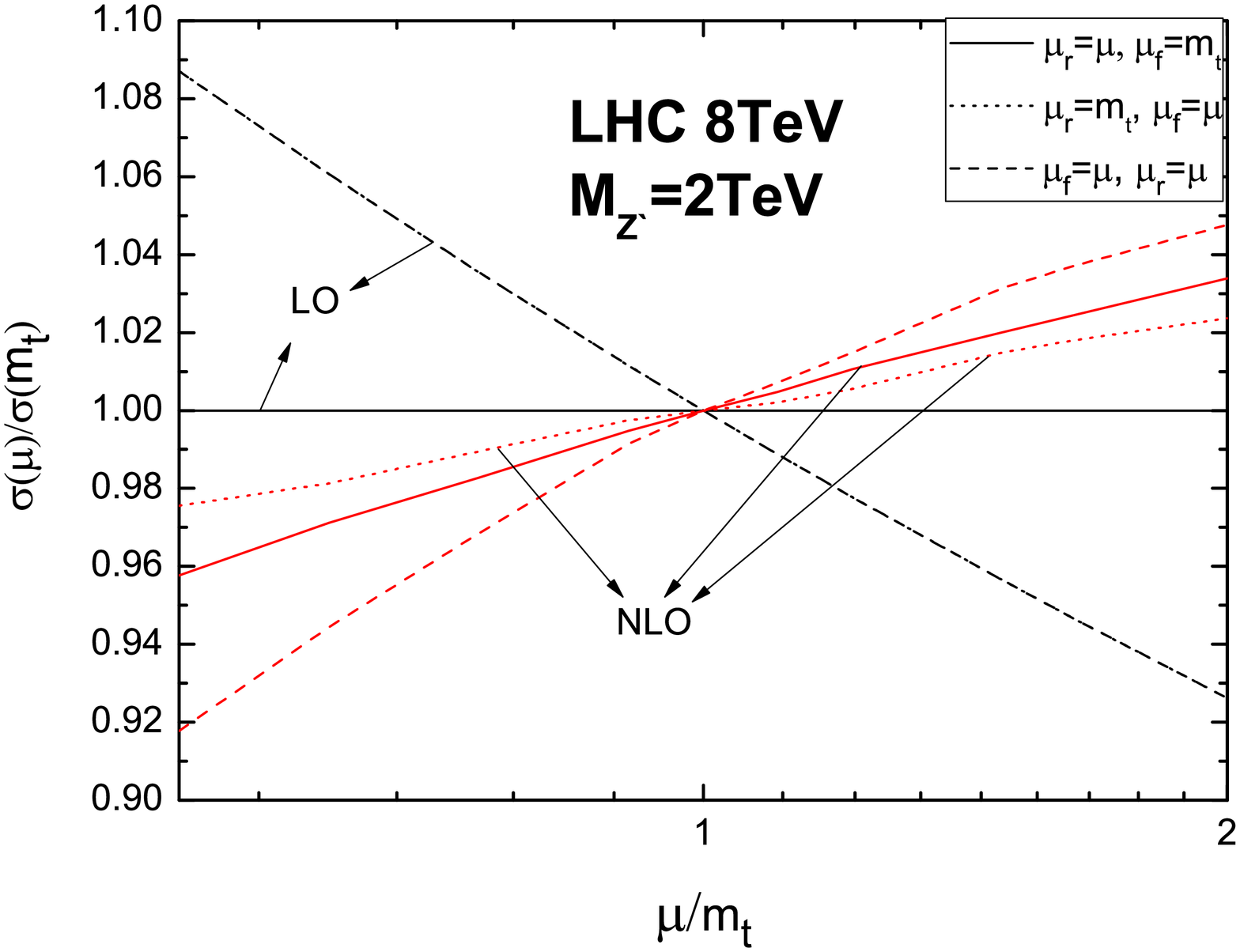}}
      \end{center}
    \end{minipage}}
          \subfigure{
    \begin{minipage}[b]{0.3\textwidth}
      \begin{center}
     \scalebox{0.201}{\includegraphics*{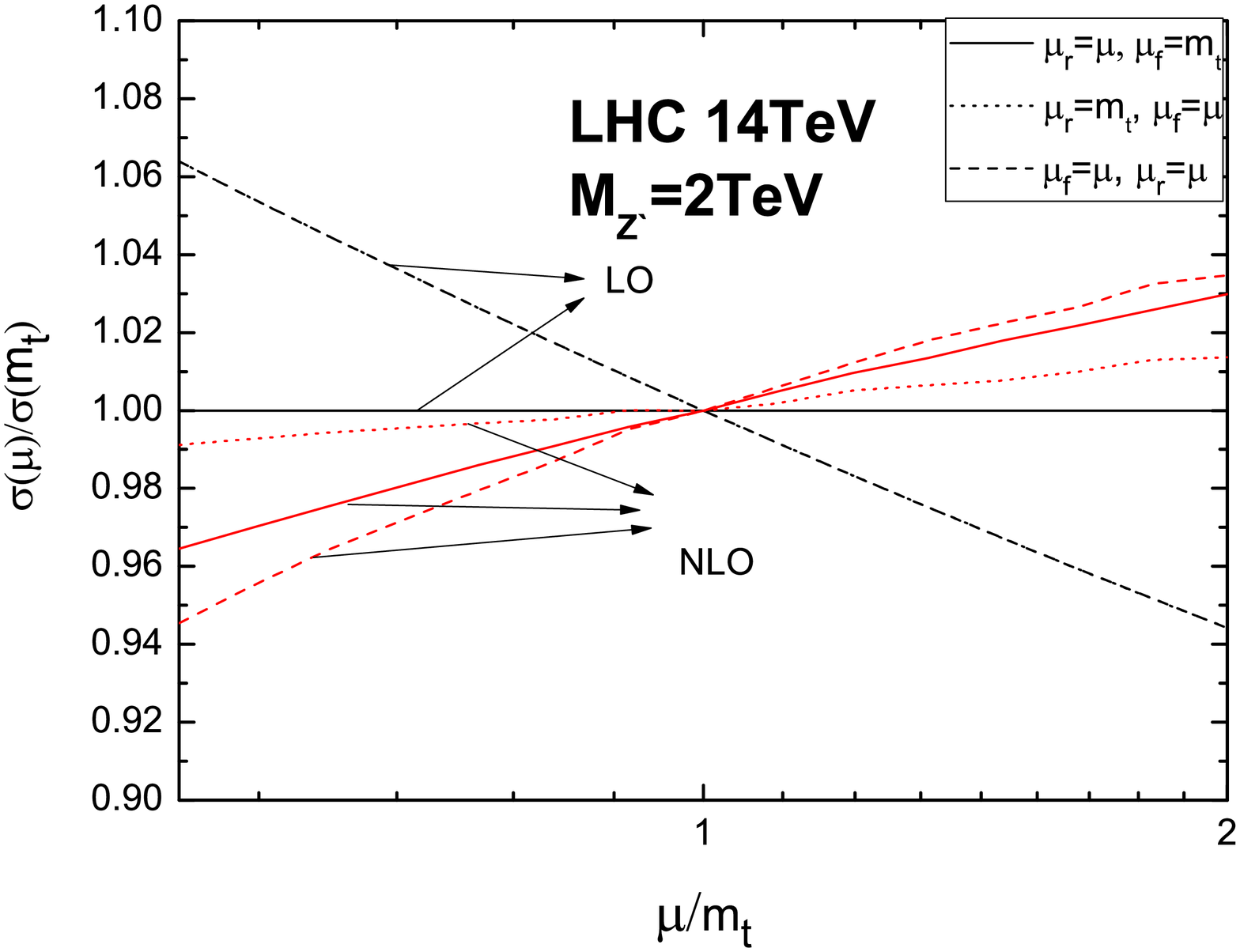}}
      \end{center}
    \end{minipage}}
    \caption{\label{scale} Scale dependence of the total cross sections at the LHC with $C_R=1$.
    The black line represents the LO result, while the red one represents the NLO result. }
\end{figure}

In Fig.~\ref{scale} we show the scale dependence of the
LO and NLO total cross section at the LHC with different parameter values.
We can see that, in all these cases,
when the scale($\mu$) varies from 0.5$m_t$ to 2$m_t$, the NLO
corrections reduce the factorization scale($\mu_f$) dependence significantly.
And at the NLO level, the total scale dependence ($\mu_f=\mu_r=\mu$) will increase when $M_Z^\prime$ increase
or $E_{CM}$ decrease.
Since there is no renormalization scale($\mu_r$) dependence at LO, the
scale dependence is just the facotrization scale dependence at LO,
which is also the reason why the total scale dependence ($\mu_f=\mu_r=\mu$)
is not reduced significantly at NLO.

\begin{figure}[H]
  \subfigure{
    \begin{minipage}[b]{0.32\textwidth}
      \begin{center}
     \scalebox{0.201}{\includegraphics*{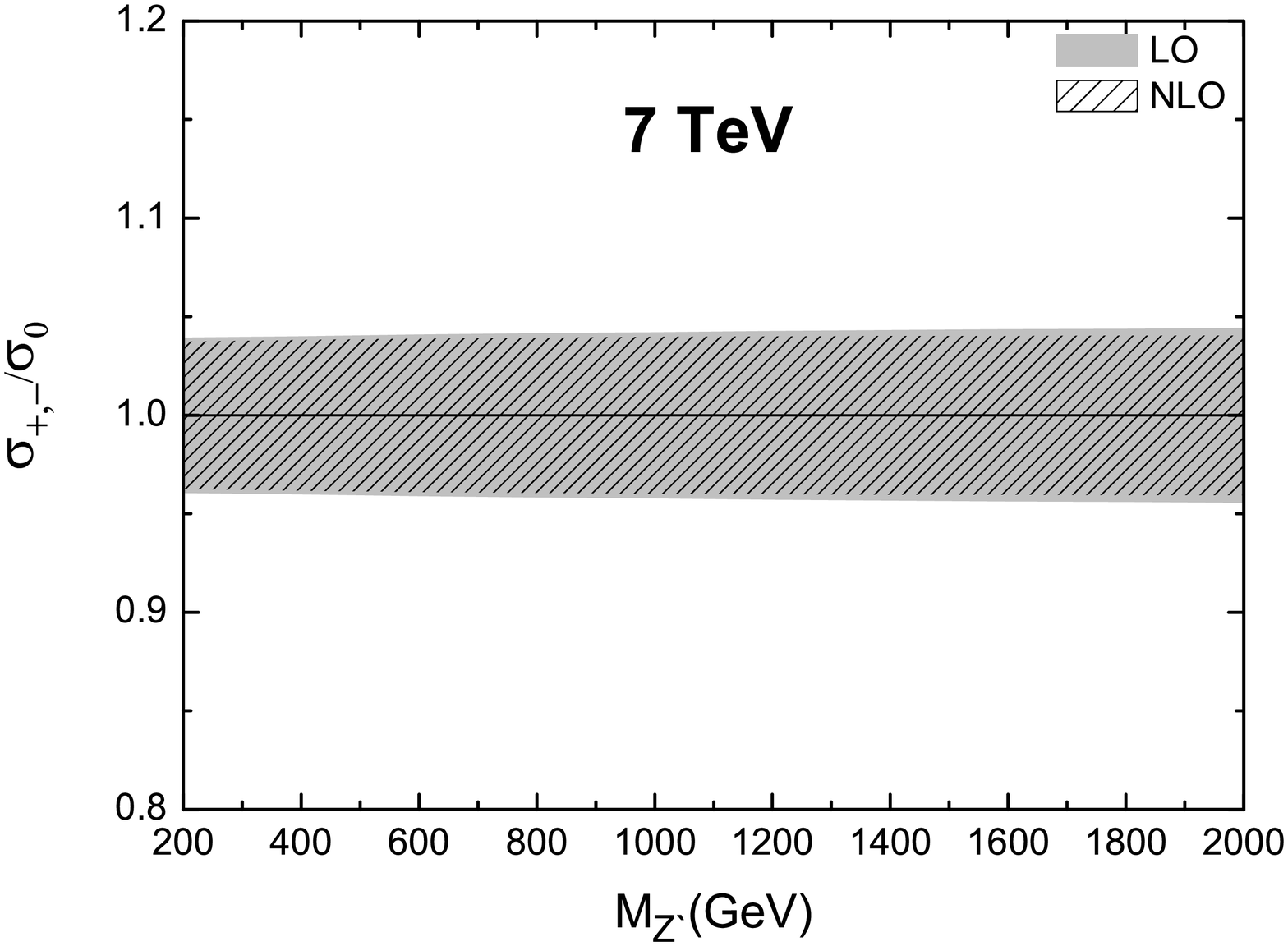}}
      \end{center}
    \end{minipage}}
  \subfigure{
    \begin{minipage}[b]{0.32\textwidth}
      \begin{center}
     \scalebox{0.201}{\includegraphics*{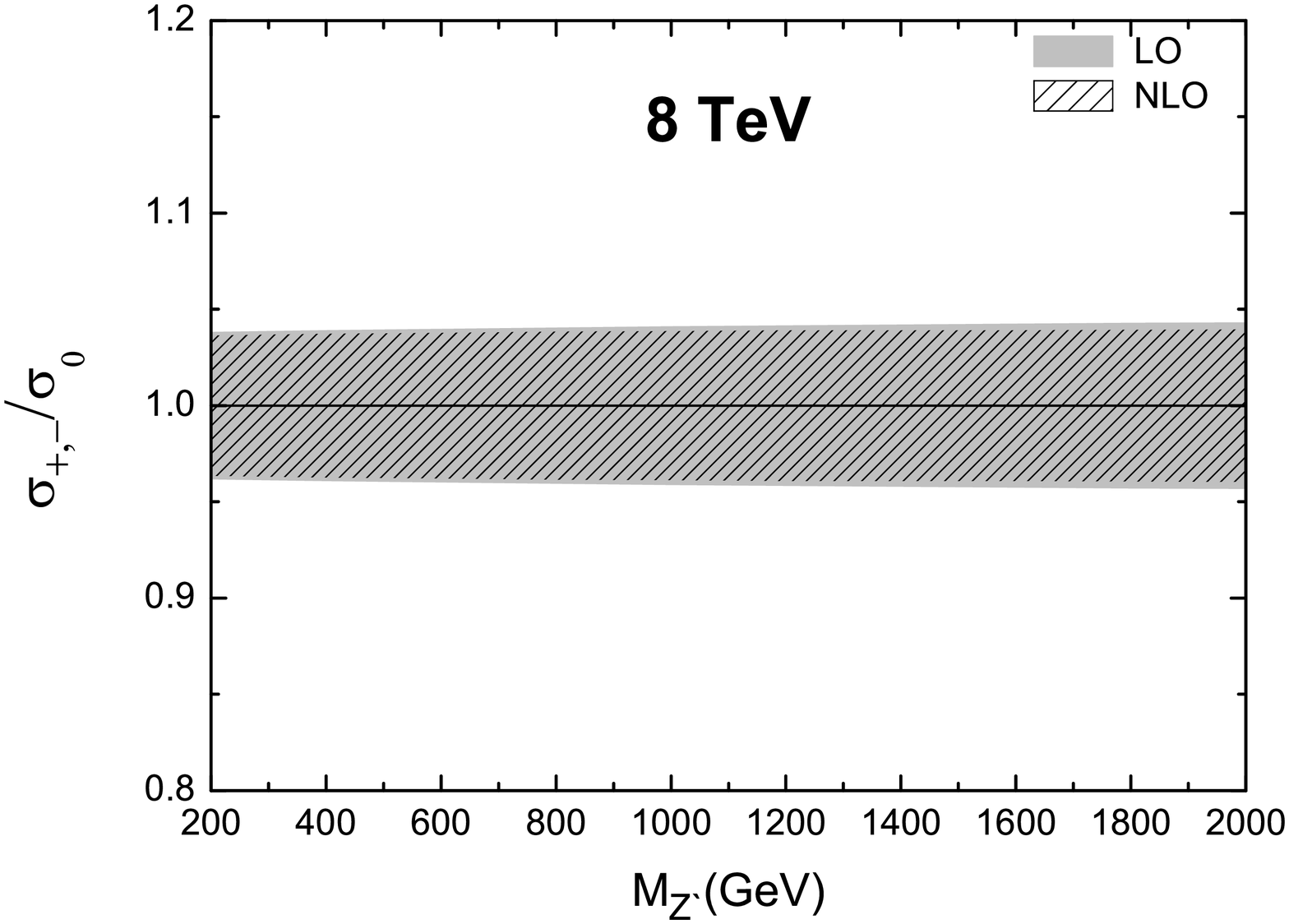}}
      \end{center}
    \end{minipage}}
      \subfigure{
    \begin{minipage}[b]{0.32\textwidth}
      \begin{center}
     \scalebox{0.201}{\includegraphics*{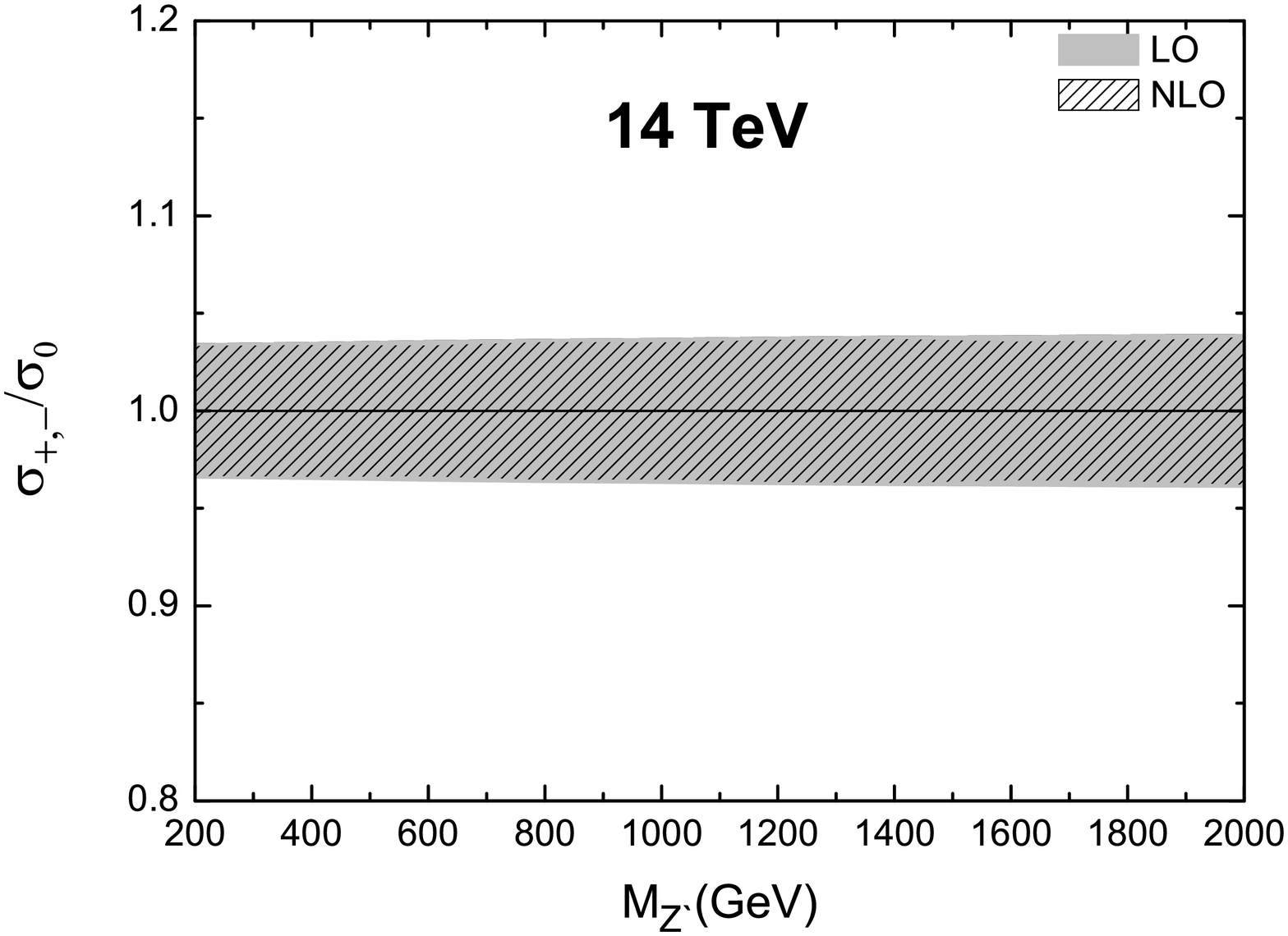}}
      \end{center}
    \end{minipage}}
    \caption{\label{pdf} PDF uncertainties with different $E_{CM}$ and $Z^{\prime}$ mass  at the LHC with $C_R=1$. $\sigma_+$  and $\sigma_-$
    are the upper and lower limits of the total cross sections respectively. $\sigma_0$ is the central value of the total cross sections.}
\end{figure}

We also discuss the  uncertainties from the PDF~\cite{Watt:2011kp} as shown in Fig.~\ref{pdf}.
According to our results, the PDF uncertainties are $3\%-4\%$, which are almost independent of
$E_{CM}$ and $M_{Z^{\prime}}$.

After including the scale uncertainty(running from 0.5$m_t$ to 2$m_t$) and the PDF uncertainties, some typical numerical results of the LO and NLO
total cross sections are shown in Table \ref{t1}, assuming $C_R=1$. It can be seen that the QCD NLO corrections can reduce the theoretical uncertainties.
\begin{table}[h]
\begin{center}
\begin{tabular}{c|c|c|c}
  \hline
    \hline
       & 7TeV & 8TeV & 14TeV  \\
   \hline
  LO ($M_{Z^{\prime}}=1TeV$)  & $179.58^{+14.28}_{-12.26}\pm7.48$	& $222.3^{+16.27}_{-14.08}\pm9.04$	& $480.58^{+22.98}_{-20.69}\pm17.75$
 \\
 NLO ($M_{Z^{\prime}}=1TeV$)  & $145.03^{+3.51}_{-6.45}\pm5.98$	& $180.87^{+4.05}_{-7.38}\pm7.27$	& $412.32^{+4.12}_{-10.57}\pm14.68$
 \\
  \hline
 LO ($M_{Z^{\prime}}=2TeV$)   & $18.86^{+1.75}_{-1.47}\pm0.83$	& $24.29^{+2.11}_{-1.79}\pm1.04$	& $60.82^{+3.88}_{-3.39}\pm2.37$
 \\
 NLO ($M_{Z^{\prime}}=2TeV$)   & $13.41^{+0.69}_{-1.13}\pm0.58$	& $17.36^{+0.83}_{-1.42}\pm0.73$	& $45.34^{+1.57}_{-2.47}\pm1.73$
 \\
  \hline
    \hline
\end{tabular}
\end{center}
\caption{The LO and the NLO total total cross sections (in fb) for the same-sign top pair production mediated by a
 nonuniversal $Z^\prime$ with different $E_{CM}$ and $Z^{\prime}$ mass.  Here the first and second errors denote the scale and PDF uncertainties, respectively.} \label{t1}
\end{table}

\subsection{Signature and background}

\begin{figure}[H]
  \subfigure{
    \begin{minipage}[b]{0.5\textwidth}
      \begin{center}
     \scalebox{0.30}{\includegraphics*{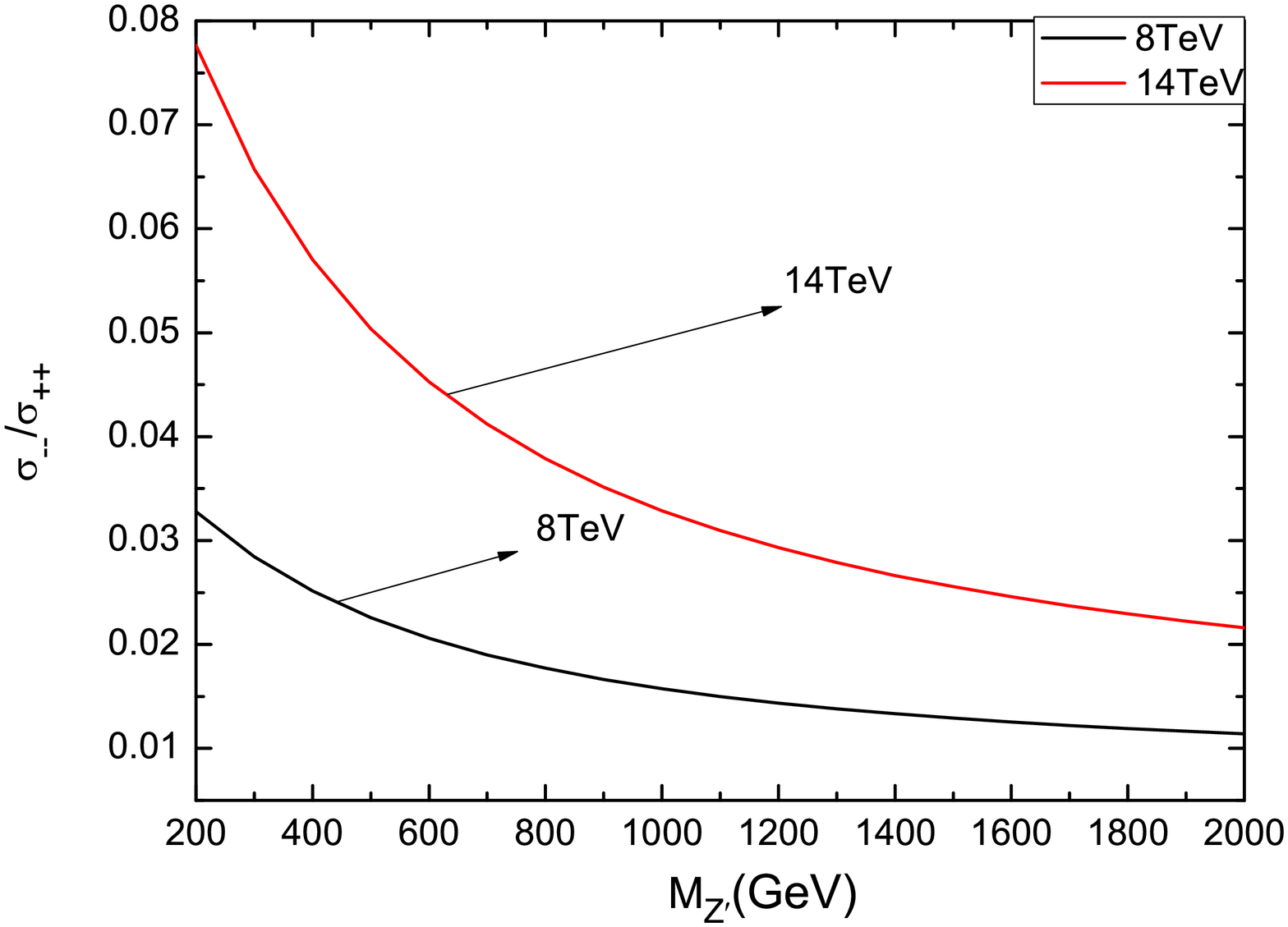}}
      \end{center}
    \end{minipage}}
  \subfigure{
    \begin{minipage}[b]{0.5\textwidth}
      \begin{center}
     \scalebox{0.30}{\includegraphics*{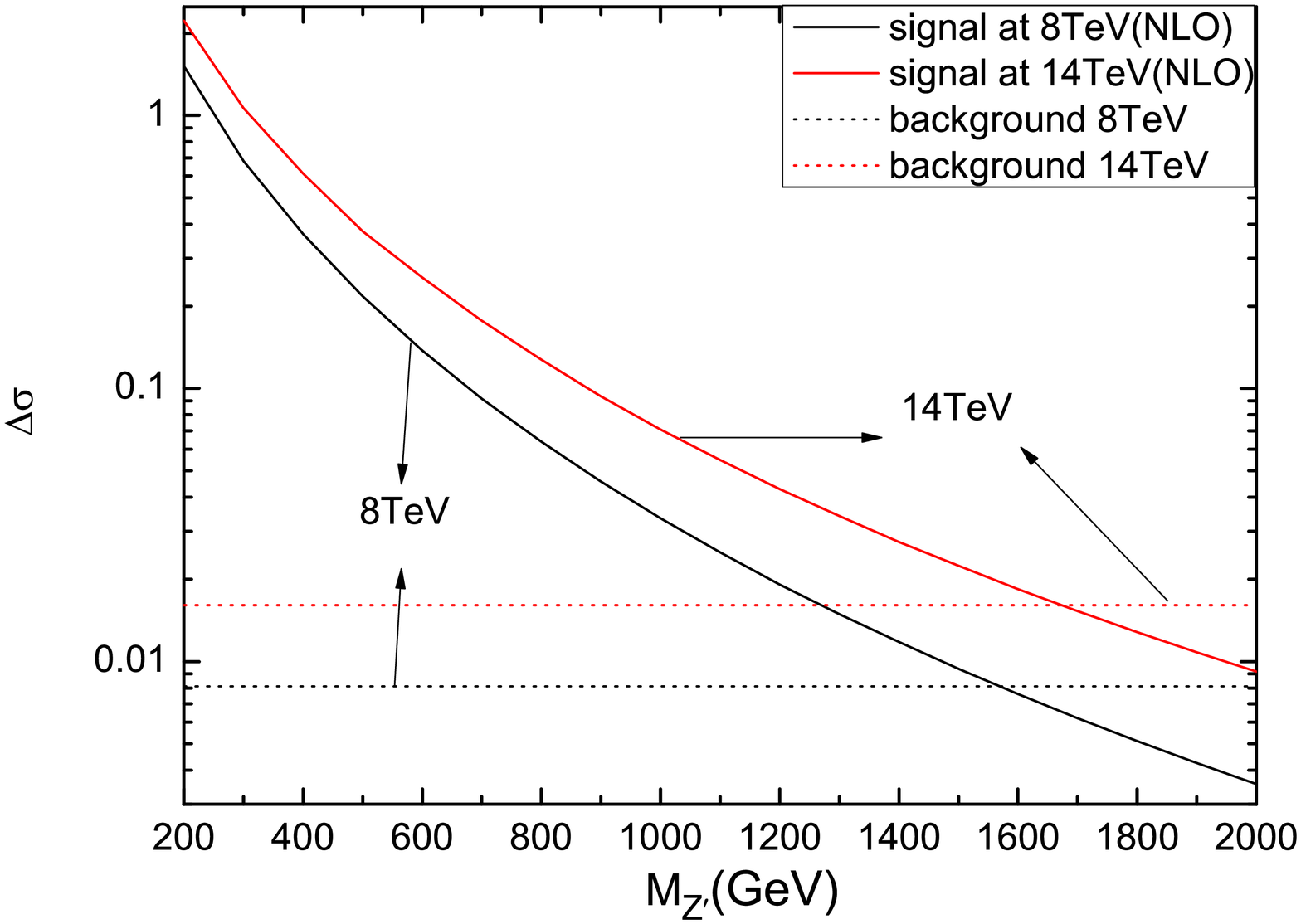}}
      \end{center}
    \end{minipage}}
    \caption{\label{ppdmm}  Ratio between $\sigma_{--}$ and $\sigma_{++}$ and $\Delta\sigma$ as a function of
     $M_Z^\prime$ at the LHC with $E_{CM}=8$~TeV
    and $E_{CM}=14$~TeV and $C_R=1$ for the same-sign top pair production process.}
\end{figure}

\begin{figure}[H]
  \subfigure{
    \begin{minipage}[b]{0.5\textwidth}
      \begin{center}
     \scalebox{0.326}{\includegraphics*{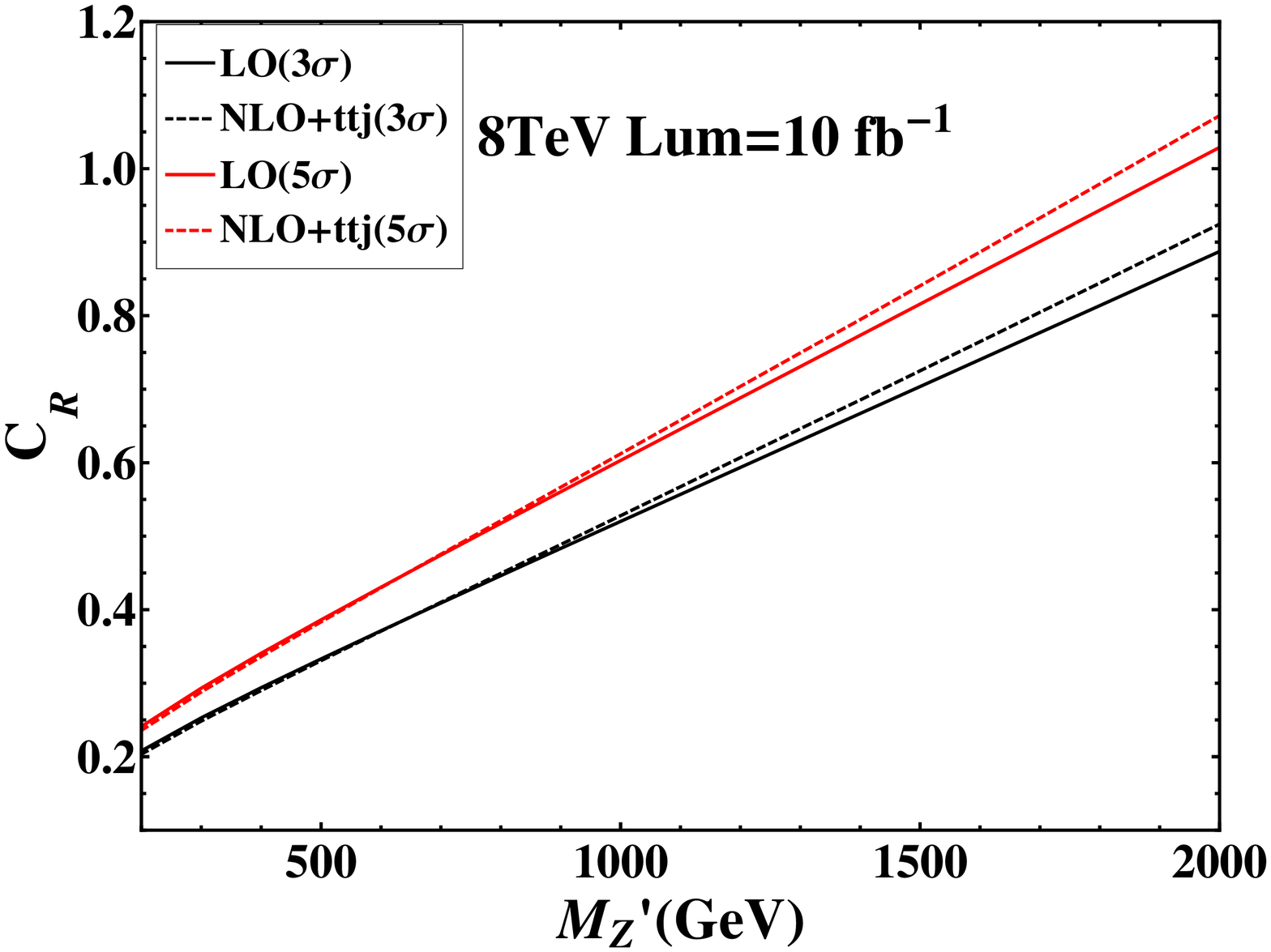}}
      \end{center}
    \end{minipage}}
  \subfigure{
    \begin{minipage}[b]{0.5\textwidth}
      \begin{center}
     \scalebox{0.32}{\includegraphics*{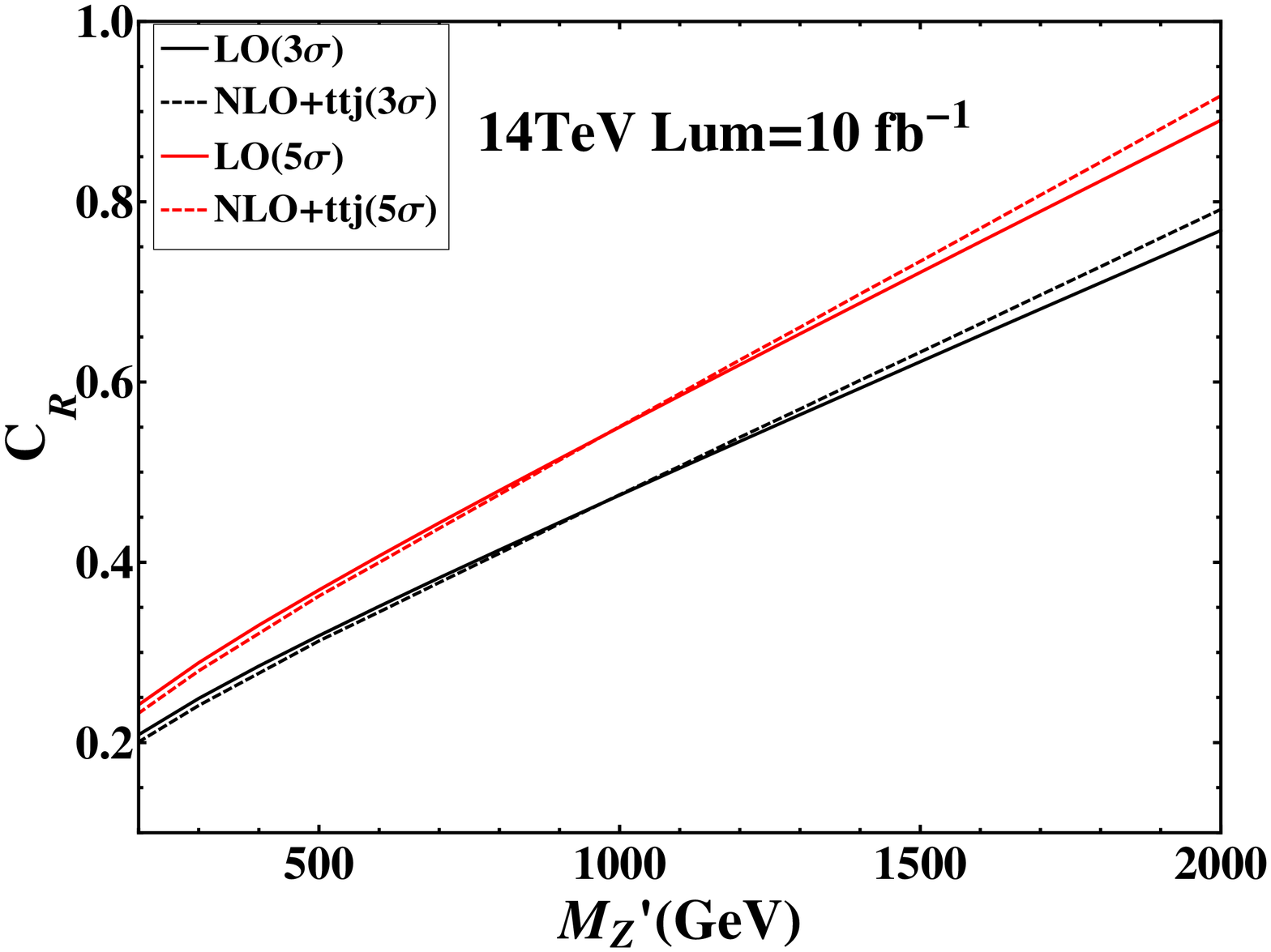}}
      \end{center}
    \end{minipage}}
    \caption{\label{ifb} The 5$\sigma$ (upper two lines) and 3$\sigma$ (lower two lines) discovery limits on $C_R$ and $M_{Z^\prime}$. }
\end{figure}

The cleanest decay mode of the same-sign top pair is to same-sign leptons.
The main background to this signal comes from the $t \bar{t}$ (when the events are selected without b-tagging), $t \bar{t}+Z$ and $t \bar{t}+W$ productions, as discussed in Refs.~\cite{Chatrchyan:2011dk, Chatrchyan:2012sa, Chatrchyan:2011wba}. In order to suppress this background, we investigate the difference between the production rate of the positively and negatively charged dilepton
\begin{equation}\label{ob}
\Delta\sigma=\sigma_{++}-\sigma_{--},
\end{equation}
where $\sigma_{++}$ and $\sigma_{--}$ are the total cross sections of the positively and negatively charged same-sign dilepton production processes, respectively. The negative same-sign leptons can be produced through the process,
$\bar{u} \bar{u} \longrightarrow \bar{t} \ \bar{t} \longrightarrow l^- \nu \bar{b} \l^- \nu \bar{b}$.
Since the LHC is a proton-proton collider, for the same-sign top pair production process, the $\sigma_{--}$
is much smaller than the $\sigma_{++}$ as shown in Fig.\ref{ppdmm}(left) because of the difference of the densities between $u$ and $\bar{u}$.
However, for the $t \bar{t}$ and $t \bar{t}+Z$ backgrounds, in principle, the $\Delta\sigma$ should vanish.
Thus, the dominant SM backgrounds are $pp \longrightarrow Wt\bar{t}$,
$pp \longrightarrow WWqq$~($q=u,d,c,s$), and $pp \longrightarrow WZqq$ when one lepton is undetected~\cite{Gao:2008vv}. But
from the Fig.\ref{ppdmm}(right), these backgrounds are also strongly suppressed (about 50$\%$) for the $\Delta\sigma$. Besides, to further suppress these backgrounds, the double b-tag and additional cuts $m_{jj}<60~{\rm GeV}$ or $m_{jj}>100~{\rm GeV}$ are required. And we give some typical numerical results of the $\Delta\sigma$
for the signal (with $M_Z^\prime=1~\text{TeV}$ and $C_R=1$) and the backgrounds in Table \ref{t2}.
We also show the 5$\sigma$ and 3$\sigma$ discovery limits on $C_R$ and $M_{Z^\prime}$ in Fig.\ref{ifb} and set $\Gamma_{Z^\prime}=\Gamma_{Z^\prime \to u\bar{t}, \bar{u}t}$ with $C_R=2$ here.
We can see that, when going from 8 to 14 TeV at LHC,
the detection capability of the 8 TeV LHC with 10$\text{fb}^{-1}$ is already good enough to study the same-sign top pair process. This is caused by the following: (1) The background and the signal increase almost at the same rates. (2) Since the cross sections are proportional to $C_R^{4}$ for a given value of $M_{Z^\prime}$, the constraints on the $C_R$ are not strong by the cross section.

\begin{table}
\begin{center}
\begin{tabular}{|c|c|c|c|c|c|c|}
\hline Process & signal(NLO) & $
Wt\bar t$ & $WWqq$ & $WZqq$
\\
\hline $ E_{CM}=8~\text{TeV} $     &32.18  &0.0064 & 0.0003 &0.0014
\\
\hline $ E_{CM}=14~\text{TeV} $    &69.03  &0.0131 & 0.0005 &0.0025
\\
\hline
\end{tabular}
\end{center}
\caption{$\Delta\sigma$ of the signal and backgrounds (in fb) after all the
cuts, assuming $M_Z^\prime=1\text{TeV}$ and $C_R = 1$.} \label{t2}
\end{table}

The observable defined in Eq.~(\ref{ob}) can also be used in other same-sign top pair process, such as those discussed in Refs.
\cite{Berger:2011ua,Gupta:2010wx,Larios:2003jq,BarShalom:2007pw,Gao:2008vv,Martin:2008aw,Zhang:2010kr}.

\begin{figure}[H]
  \subfigure{
    \begin{minipage}[b]{0.5\textwidth}
      \begin{center}
     \scalebox{0.32}{\includegraphics*{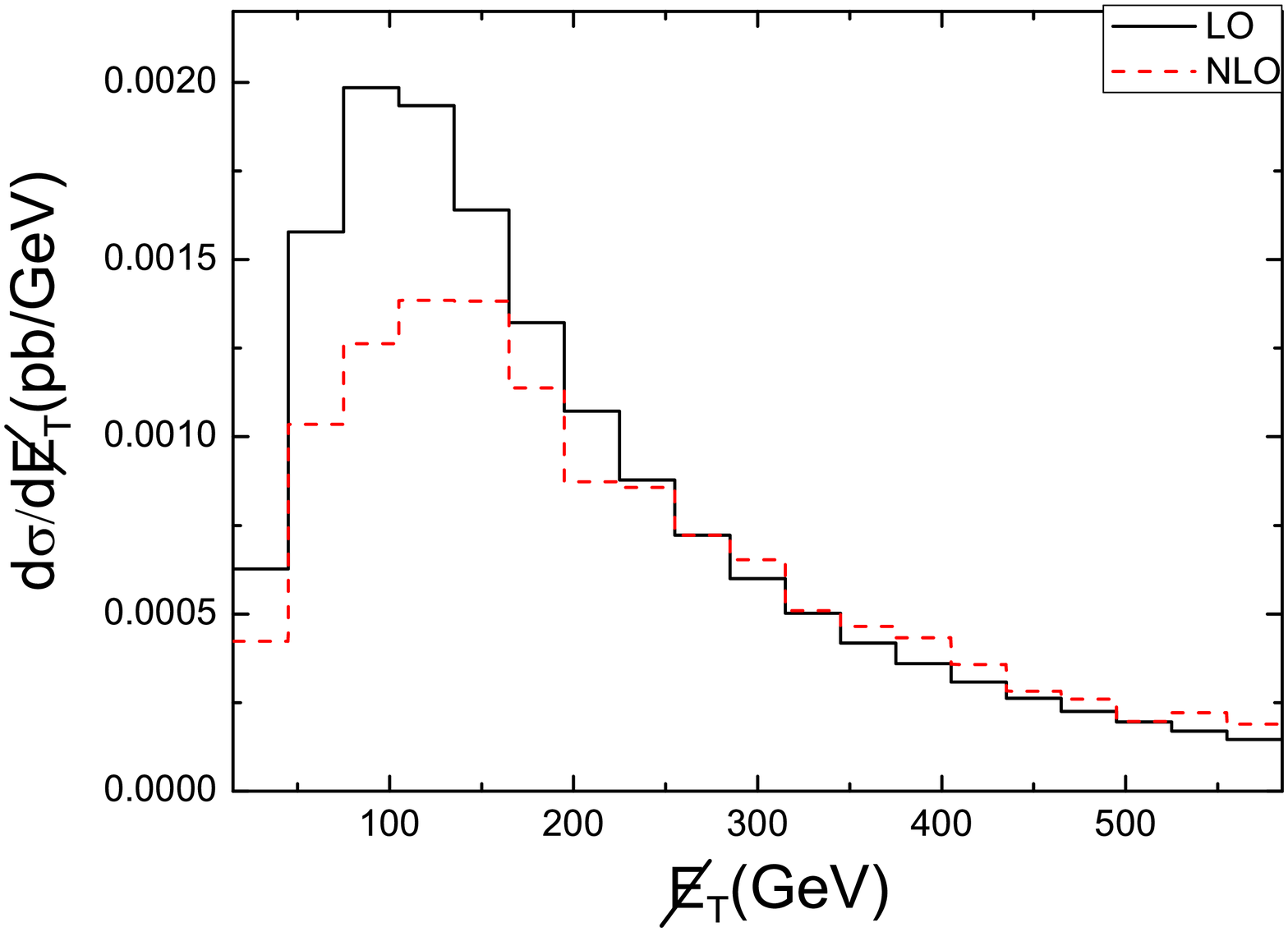}}
      \end{center}
    \end{minipage}}
  \subfigure{
    \begin{minipage}[b]{0.5\textwidth}
      \begin{center}
     \scalebox{0.32}{\includegraphics*{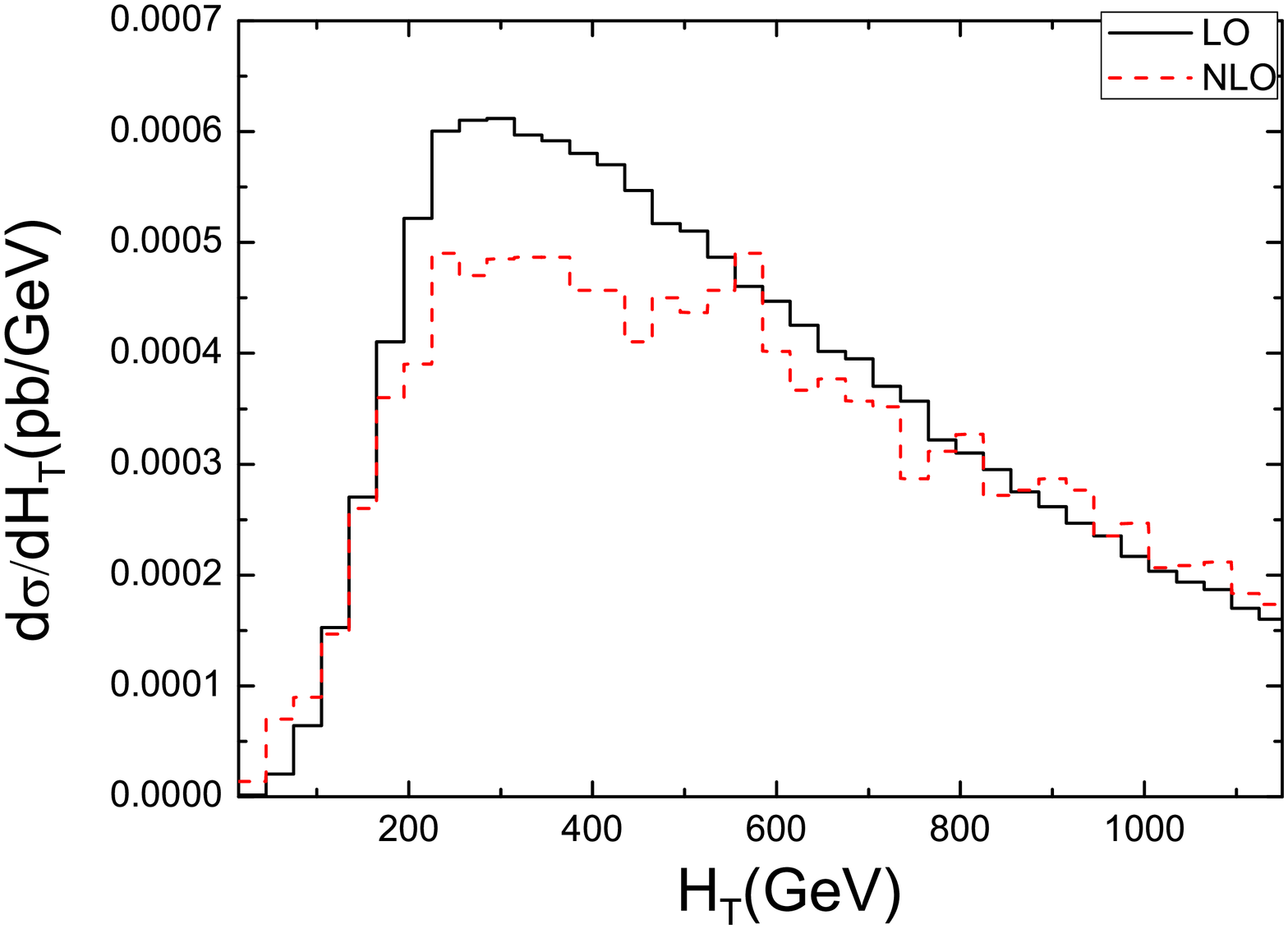}}
      \end{center}
    \end{minipage}}
    \caption{\label{kim} $\slashed E_T$ and $H_T$ distributions at the LHC with $\sqrt{S}=14$~TeV. Here, we assume $C_R=1$ and $M_{Z^\prime}=1$~TeV.}
\end{figure}

In Fig.~\ref{kim} we show the dependence of the differential distributions on missing energy~($\slashed E_T$) and $H_T$ (scalar sum of final-state visible particle transverse momenta). From Fig.~\ref{kim} we can see
that QCD NLO corrections reduce significantly the distributions of missing energy and $H_T$ in the ranges of 70-200~GeV and 200-600~GeV, respectively. This is due to the fact that the real corrections have additional jet emissions.

\section{conclusions}\label{s6}

In conclusion we have investigated the complete QCD NLO corrections to the
same-sign top pair production mediated by the nonuniversal $Z^\prime$ including production
and decay at the LHC. Our results show that the QCD NLO corrections reduce the total cross sections by more than 10\% for $Z^\prime$ boson mass greater than 500 GeV and loosen the constraint on the model parameters when comparing with the CMS Collaboration results. We also show that the total cross sections, including the QCD NLO corrections, can be expressed as the explicit functions of the model parameters $C_R$ and $M_{Z^\prime}$ generally. These functions may help experimentalists to quickly estimate the cross sections in their studies.
Besides, the NLO corrections reduce the dependence of the total cross sections on the factorization scale significantly. We also study the signature and backgrounds of the process at the NLO level.
Using the difference between the $\sigma_{++}$ and the $\sigma_{--}$ as shown in Eq.(\ref{ob}), we show that, in principle, the  $t\bar{t}$ and $t\bar{t}+Z$ backgrounds can be totally excluded, and the same-sign dilepton signal of the new physics could be discovered more easily.

\begin{acknowledgments}
This work was supported in part by National Nature Science Foundation of China,
under Grants No. 11021092, No. 10975004 and No. 11135003.
\end{acknowledgments}


\end{document}